\documentclass[%
 reprint,
 amsmath,amssymb,
 aps,
]{revtex4-2}

\usepackage{graphicx}
\usepackage{dcolumn}
\usepackage{bm}
\usepackage{hyperref}
\usepackage{times}

\hypersetup{
    unicode = true,           
    pdftoolbar = false,        
    pdfmenubar = false,        
    pdffitwindow = false,      
    pdfstartview = {FitH},     
    colorlinks = true,         
    citecolor = blue,           
    linkcolor = blue,
    urlcolor = blue
}

\DeclareMathOperator{\csch}{csch}

\begin{document}

\preprint{APS/123-QED}

\title{Nanoscale Spin Injector Driven by a Microwave Voltage}

\author{A. I. Nikitchenko }
\author{N. A. Pertsev }
 \email{pertsev.domain@mail.ioffe.ru}
\affiliation{%
 Ioffe Institute, St. Petersburg 194021, Russia\\
}%





\begin{abstract}
We propose an electrically driven spin injector into normal metals and semiconductors, which is based on a magnetic tunnel junction (MTJ) subjected to a microwave voltage. Efficient functioning of such an injector is provided by electrically induced magnetization precession in the ``free'' layer of MTJ, which generates the spin pumping into a metallic or semiconducting overlayer. To validate the feasibility of the proposed device, we theoretically describe the spin and charge dynamics in the CoFeB/MgO/CoFeB/Au and CoFeB/MgO/CoFeB/GaAs tunneling heterostructures. First, the magnetization dynamics in the free CoFeB layer is quantified with the account of a spin-transfer torque generated by the spin-polarized current flowing through the MTJ and a voltage-controlled magnetic anisotropy associated with the CoFeB$|$MgO interface. The calculations are performed in the macrospin approximation for an ultrathin CoFeB layer with perpendicular anisotropy and nanoscale in-plane dimensions. By numerically solving the Landau-Lifshitz-Gilbert-Slonczewski equation, we determine dependences of the precession amplitude on the frequency $f$ and magnitude $V_\mathrm{max}$ of the ac voltage applied to the MTJ. It is found that the frequency dependence changes drastically above the threshold amplitude $V_\mathrm{max} \approx 200$~mV, exhibiting a break at the resonance frequency $f_\mathrm{res}$ due to nonlinear effects. The results obtained for the magnetization dynamics are then used to describe the spin injection and pumping into the Au and GaAs overlayers. The total spin-current density near the interface is calculated as a function of time at different excitation frequencies and voltage amplitudes. Since the generated spin current creates additional charge current owing to the inverse spin Hall effect, we also calculate distributions of the charge-current density and electric potential in the thick Au overlayer. The calculations show that the arising transverse voltage, which can be used to probe the efficiency of spin generation electrically, becomes experimentally measurable at $f = f_\mathrm{res}$. Finally, we evaluate the spin accumulation in a long n$^+$-GaAs bar coupled to the MTJ and determine its temporal variation and spatial distribution along the bar. It is found that the ac spin accumulation under resonant excitation is large enough for experimental detection via a voltage between two ferromagnetic nanocontacts even at micrometer distances from the MTJ. This result demonstrates high efficiency of the described nanoscale spin injector driven by microwave voltage.

\end{abstract}

\maketitle


\section{\label{sec:intro}INTRODUCTION}

Efficient spin injectors are necessary for the functioning of various spintronic devices, such as spin diodes, spin field-effect transistors, magnetic bipolar transistors, hot-electron spin transistors, and spin-based logic gates~\cite{Zutic2004,Dery2007}. The key element of such injectors is a ferromagnet providing a spin imbalance in a neighboring normal metal or semiconductor. In all-metallic heterostructures, the spin imbalance can be achieved by simply injecting a spin-polarized current from the ferromagnetic emitter into a paramagnetic or diamagnetic metal~\cite{JohnsonSilsbee1985,Jedema2001}. However, this direct spin injection becomes inefficient for semiconductors due to an impedance mismatch at the interface~\cite{Schmidt2000}. The problem can be solved by inserting of a tunnel barrier at a ferromagnet-semiconductor interface, as predicted theoretically~\cite{Rashba2000} and demonstrated experimentally at room temperature for a CoFe-MgO tunnel injector into GaAs~\cite{Jiang2005}. However, the injected spin-polarized carriers are hot, whereas cold-electron spin injection is desirable for semiconductor devices~\cite{Tserkovnyak2005}.

Another method to create a spin current and spin accumulation in a normal conductor is based on the spin pumping generated by a ferromagnet with precessing magnetization~\cite{Tserkovnyak2005}. When the precessing metallic ferromagnet is brought into Ohmic contact with the conductor, it represents a ``spin battery'', which becomes effective at the ferromagnetic resonance~\cite{Brataas2002}. Efficient spin pumping into various normal metals has been achieved with the aid of metallic and insulating ferromagnetic films excited by microwave magnetic fields~\cite{Heinrich2003, Saitoh2006, Mosendz2010PRB, Sandweg2010, Ando2011JAP}. Room-temperature generation of spin flow in semiconductors through both Ohmic and Schottky contacts was demonstrated by this technique as well~\cite{Ando2011NM, Shikoh2013}. However, the use of microwave magnetic fields has serious disadvantages for practical applications, such as associated high energy losses and issues related with the downscaling of spintronic devices. Fortunately, the magnetization precession can also be excited electrically using spin-polarized currents~\cite{Kiselev2003, Tulapurkar2005, Sankey2006, Deac2008}, electric-field-dependent magnetic anisotropy~\cite{Nozaki2012, Zhu2012, Viaud2014, Miura2017}, and piezoelectrically generated elastic waves~\cite{Weiler2011, Weiler2012, Azovtsev2016, Polzikova2018, Azovtsev2019}, which opens the possibility to develop spin injectors with greatly reduced power consumption.

In this paper, we theoretically study a nanoscale magnetic tunnel junction (MTJ) subjected to a microwave voltage and show that it can be employed as an efficient spin injector into normal metals and semiconductors. To this end, we first describe electrically induced magnetization precession in the ``free'' layer of a CoFeB/MgO/CoFeB junction with the account of additional damping caused by the spin pumping into a metallic or semiconducting overlayer~\cite{Tserkovnyak2002}. The calculations are carried out in the macrospin approximation via numerical integration of the Landau-Lifshitz-Gilbert-Slonczewski (LLGS) equation. Since the spin pumping generated by the precession should intensify with increasing amplitude $V_\mathrm{max}$ of the ac voltage applied to the MTJ, our analysis focuses on peculiarities of the magnetization dynamics appearing in the range of enhanced amplitudes $V_\mathrm{max}$, where nonlinear effects become important. The results obtained for the magnetization precession in the free CoFeB layer are then employed to calculate the spin injection and pumping into an Au overlayer and into a GaAs bar coupled to the MTJ. To evaluate the efficiency of the proposed spin injector, which can be probed electrically~\cite{Saitoh2006, JohnsonSilsbee1985}, we also quantify the charge flow and the distribution of electric potential in the Au overlayer and calculate the spin accumulation in the GaAs bar.

\section{\label{sec:dynamics}MAGNETIZATION DYNAMICS\\ DRIVEN BY MICROWAVE VOLTAGE}

Magnetic dynamics in MTJs can be generated electrically because a spin-polarized current creates a spin-transfer torque (STT) when the magnetizations of two electrodes are noncollinear~\cite{Slonczewski1989, SlonczewskiSun2007}.
\begin{figure}[b]
\centering
\includegraphics[width=1.0\linewidth]{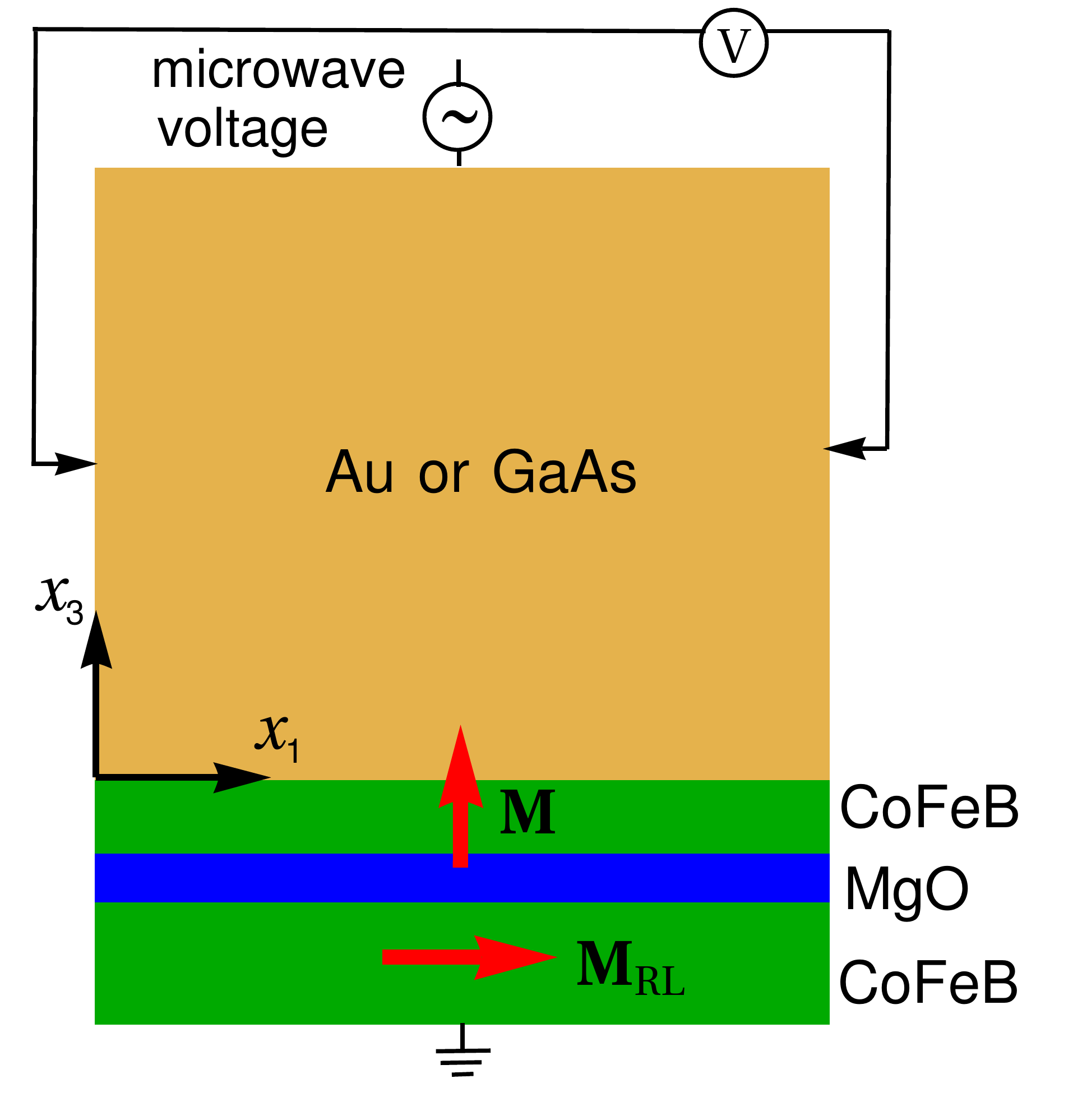}
\caption{\label{fig:mtj} Design of electrically driven spin injector based on magnetic tunnel junction subjected to a microwave voltage. The studied heterostructure comprises CoFeB/MgO/CoFeB tunnel junction with an ultrathin free layer covered by Au or GaAs. Information on the spin injection into the overlayer can be obtained by measuring the voltage $V$ caused by the inverse spin Hall effect with the aid of nanowires brought into contact with opposite lateral sides of the overlayer.}
\end{figure}
Furthermore, in MgO-based MTJs having voltage-controlled magnetic anisotropy (VCMA), magnetization oscillations can be induced in the free layer (FL) by microwave-frequency voltages even in the absence of significant STTs~\cite{Nozaki2012, Viaud2014, Miura2017}. To enhance the STT acting on the FL magnetization, one can employ an MTJ with an ultrathin FL having perpendicular magnetic anisotropy~\cite{Ikeda2010, Kanai2012} and a thick reference layer (RL) with in-plane magnetization (Fig.~\ref{fig:mtj}). This feature motivated us to consider such a geometry in our study, where we focus on the Co$_{20}$Fe$_{60}$B$_{20}$/MgO/Co$_{20}$Fe$_{60}$B$_{20}$ junction having pronounced VCMA~\cite{Zhu2012}. It should be noted that, owing to the interlayer exchange coupling (IEC) between FL and RL, the FL magnetization \textbf{M} slightly deviates from the perpendicular-to-plane orientation (polar angle $\theta \approx 3^\circ$) even in the absence of applied voltage and external magnetic field. Therefore, the VCMA contributes to the voltage-induced destabilization of the FL magnetization along with the STT~\cite{Viaud2014}.

To quantify the magnetization oscillations $\delta \mathbf{M}(t)$ generated by the applied ac voltage $V_\mathrm{ac} = V_\mathrm{max} \sin{(2 \pi f t)}$ in FL with nanoscale in-plane dimensions, we use the LLGS equation and the macrospin approximation, which implies the same magnetization direction $\mathbf{m} = \mathbf{M} / M_s$ in the whole FL and a constant saturation magnetization $M_s$. RL is assumed to be uniformly magnetized with a fixed magnetization direction $\mathbf{m}_\mathrm{RL}$ unaffected by the applied voltage, which is confirmed by numerical calculations at the considered RL thickness $t_\mathrm{RL} = 3$~nm. Since the field-like torque does not change the magnetic dynamics qualitatively~\cite{Zhu2012, Fang2016}, we write the LLGS equation in the form

\begin{equation}
    \begin{gathered}
        \frac{d\mathbf{m}}{dt} = - \gamma \mu_0 \mathbf{m} \times \mathbf{H}_\mathrm{eff} + \alpha \mathbf{m} \times \frac{d\mathbf{m}}{dt}\\ + \frac{\tau_\mathrm{STT}}{M_s} \mathbf{m} \times (\mathbf{m} \times \mathbf{m}_\mathrm{RL}),
    \end{gathered}
    \label{eq:LLGS}
\end{equation}
\noindent
where $\gamma > 0$ is the electron's gyromagnetic ratio, $\mu_0$ is the permeability of vacuum, $\alpha$ is the Gilbert dimensionless damping parameter, and $\mathbf{H}_\mathrm{eff}$ is the effective magnetic field acting on the FL magnetization. The last term in Eq. (\ref{eq:LLGS}) allows for the STT created by the spin-polarized current flowing across FL, with the factor $\tau_\mathrm{STT}$ being proportional to the applied voltage $V = V_\mathrm{dc} + V_\mathrm{ac}$ in the first approximation. In the case of elastic tunneling in symmetric MTJs, the theoretical calculations yield $\tau_\mathrm{STT} = (\gamma \hbar / 2 e)(V G_\mathrm{P} / t_\mathrm{FL}) \eta / (1 + \eta^2)$, where $e > 0$ is the elementary charge, $\hbar$ is the reduced Planck constant, $t_\mathrm{FL}$ is the FL thickness, $\eta = \sqrt{(G_\mathrm{P} - G_\mathrm{AP}) / (G_\mathrm{P} + G_\mathrm{AP})}$ is the MTJ asymmetry factor, and $G_\mathrm{P}$ and $G_\mathrm{AP}$ are the junction's conductances per unit area at parallel and antiparallel electrode magnetizations, respectively~\cite{SlonczewskiSun2007}. To take into account the influence of the precession-induced spin pumping into a metallic or semiconducting overlayer, we renormalize the parameters $\gamma$ and $\alpha$ involved in Eq.~(\ref{eq:LLGS}) as~\cite{Tserkovnyak2002}

\begin{equation}
    \begin{gathered}
    \alpha = \frac{\gamma}{\gamma_0} \bigg(\alpha_0 + \frac{g_L \mu_B}{4 \pi M_s t_\mathrm{FL}} \mathrm{Re}\big[g^r_{\uparrow \downarrow}\big]\bigg),\\
    \frac{1}{\gamma} = \frac{1}{\gamma_0}\bigg(1 + \frac{g_L \mu_B}{4 \pi M_s t_\mathrm{FL}} \mathrm{Im}\big[g^r_{\uparrow \downarrow}\big]\bigg),
    \end{gathered}
\label{eq:SPrenorm}
\end{equation}
\noindent
where $\gamma_0$ and $\alpha_0$ denote the values of $\gamma$ and $\alpha$ in the absence of spin pumping, $g_L$ is the Land\'{e} factor, $\mu_B$ is the Bohr magneton, and $g^r_{\uparrow \downarrow}$ is the complex reflection spin-mixing conductance per unit area of the FL-overlayer contact~\cite{Zwierzycki2005}. Since in our case the numerical estimates demonstrate a negligible dependence of the Gilbert parameter $\alpha_0$ on the magnetization precession power~\cite{Slavin2009}, we consider $\alpha_0$ as a constant quantity.

For a homogeneously magnetized ultrathin CoFeB layer, the field $\mathbf{H}_\mathrm{eff} = -(\mu_0 M_s)^{-1} \partial F / \partial \mathbf{m}$ involved in Eq.~(\ref{eq:LLGS}) can be determined by differentiating the effective volumetric Helmholtz free-energy density $F$ of that layer. The magnetization-dependent part $\Delta F(\mathbf{m})$ of this energy may be written as
\begin{widetext}
\begin{equation}
    \begin{gathered}
        \Delta F \simeq K_1 (m_1^2 m_2^2 + m_1^2 m_3^2 + m_2^2 m_3^2) + \frac{K_s}{t_\mathrm{FL}} m_3^2 + \frac{1}{2} \mu_0 M_s^2(N_{11} m_1^2 + N_{22} m_2^2 + N_{33} m_3^2\\
        + 2 N_{12}m_1 m_2 + 2 N_{13}m_1 m_3 + 2 N_{23}m_2 m_3) - \frac{U_\mathrm{IEC}}{t_\mathrm{FL}} \mathbf{m} \cdot \mathbf{m}_\mathrm{RL},
    \end{gathered}
    \label{eq:F}
\end{equation}
\end{widetext}
\noindent
where $m_i$ ($i = 1, 2, 3$) are the projections of \textbf{m} on the crystallographic axes $x_i$ of the CoFeB layer, which is assumed to be epitaxial with the $x_3$ axis orthogonal to its surfaces, $K_1$ characterizes the cubic magnetocrystalline anisotropy of CoFeB~\cite{Ikeda2012}, $K_s$ defines the total specific energy of the FL surfaces, $U_\mathrm{IEC}$ is the IEC energy per unit area, $N_{ij}$ are the demagnetizing factors, and the external magnetic field is absent in our case. Since the magnetic anisotropy associated with the Co$_{20}$Fe$_{60}$B$_{20}|$MgO interface varies with the electric field $E_3$ created in MgO~\cite{Kanai2012, Alzate2014}, the factor $K_s$ appears to be a voltage-dependent quantity. In the linear approximation supported by first-principles calculations~\cite{Niranjan2010} and experimental data~\cite{Alzate2014}, $K_s = K_s^0 + k_s V / t_b$, where $K_s^0 = K_s(E_3=0)$, $k_s = \partial k_s / \partial E_3$ is the electric-field sensitivity of $K_s$, and $t_b$ is the thickness of MgO tunnel barrier. 

\begin{figure*}[ht!]
\centering
\includegraphics[width=0.32\linewidth]{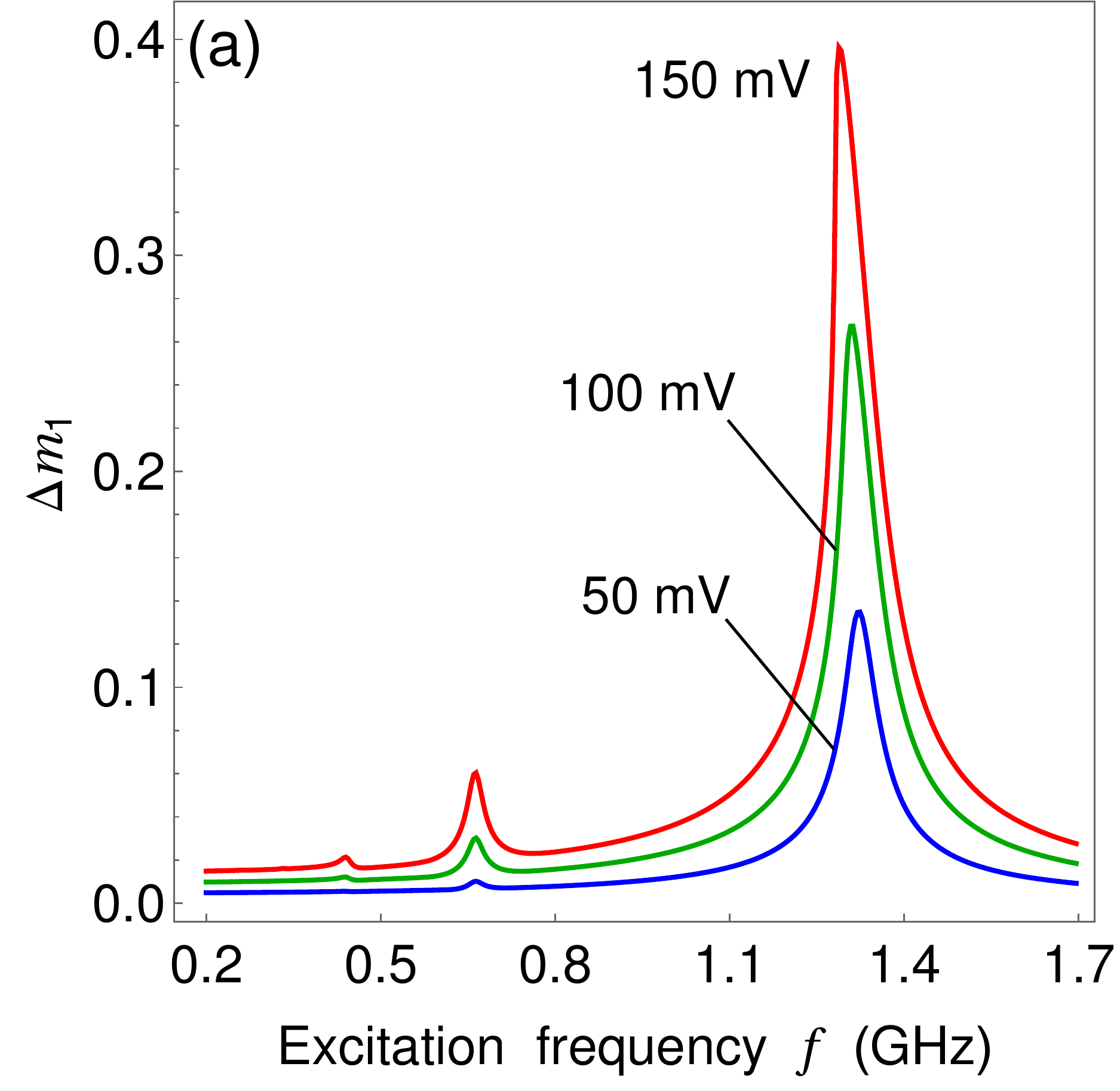}
\hfill
\includegraphics[width=0.32\linewidth]{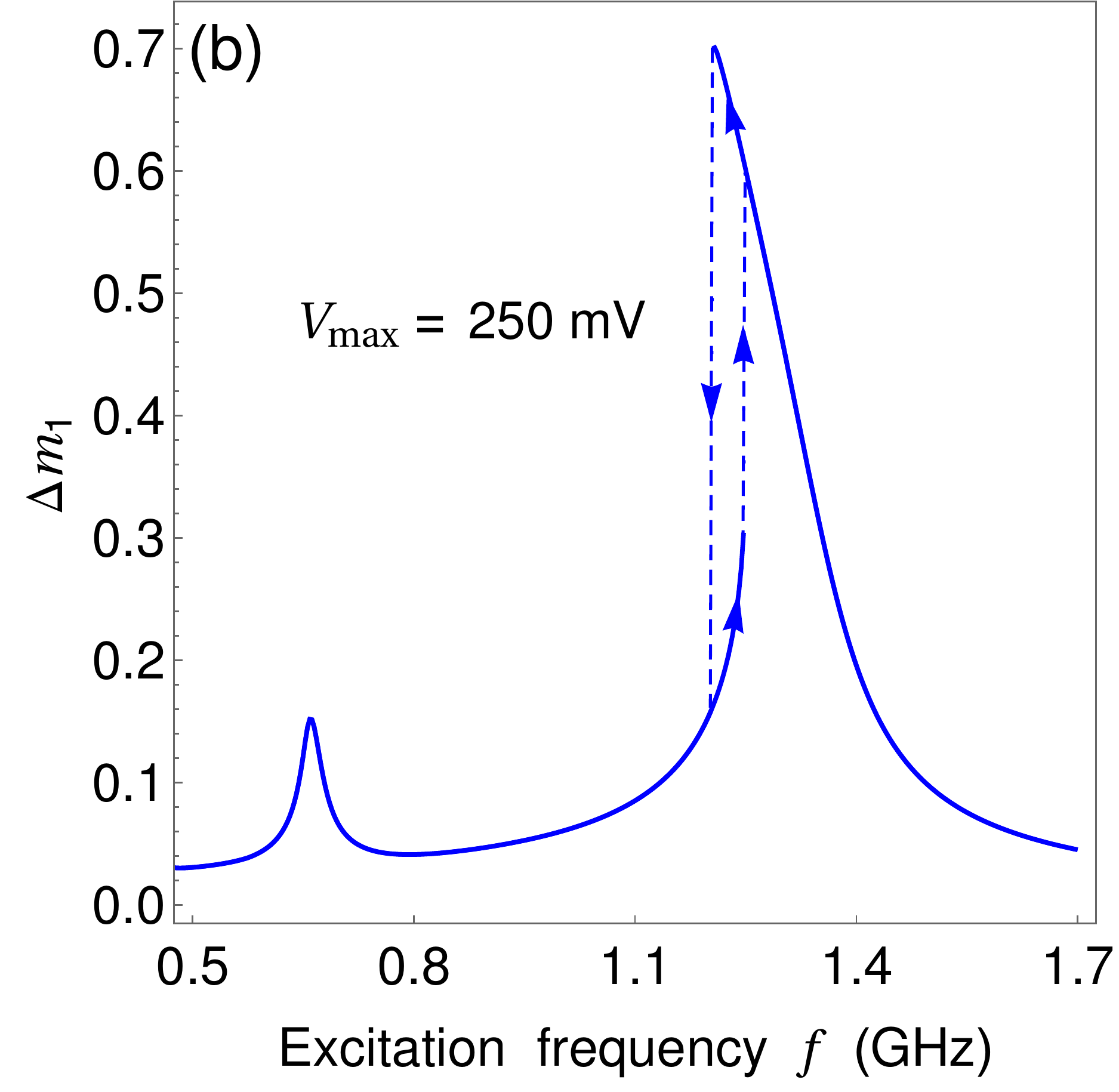}
\hfill
\includegraphics[width=0.32\linewidth]{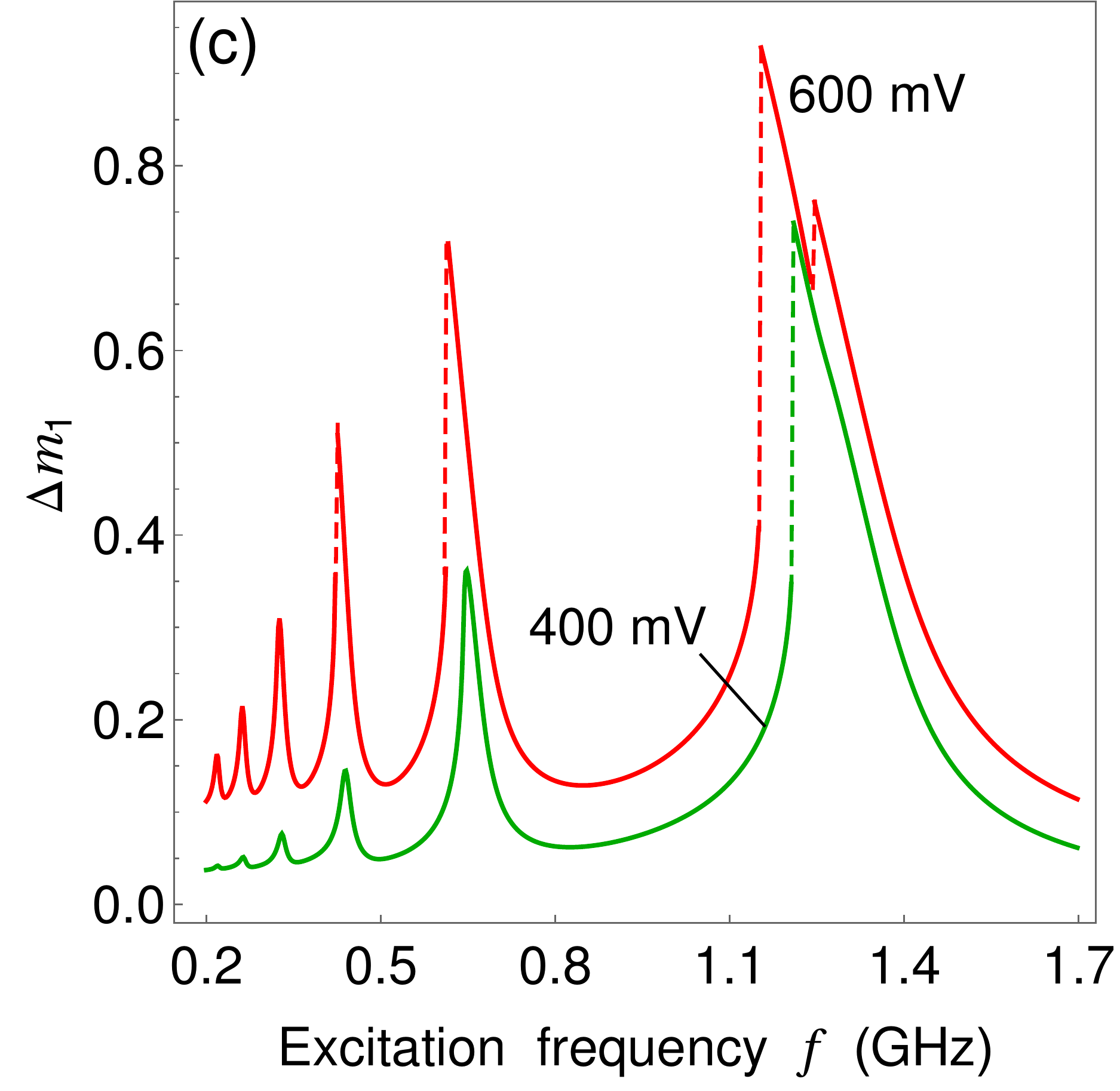}
\caption{\label{fig:res} Frequency dependences of the amplitude of magnetization precession in the free CoFeB layer covered by Au calculated at different amplitudes $V_\mathrm{max}$ of the microwave voltage applied to the tunnel junction. Each new simulation starts from a steady-state precession generated during the previous simulation performed at a slightly differing excitation frequency. The graphs show the swing $\Delta m_1$ of the scalar product $\mathbf{m} \cdot \mathbf{m}_\mathrm{RL}$ of the unit vectors directed along the FL and RL magnetizations. Voltage amplitudes $V_\mathrm{max}$ are indicated near the curves. Panels (a) and (b) present $\Delta m_1(f)$ calculated at both increasing and decreasing voltage frequency $f$, whereas panel (c) shows results only for increasing frequency.}
\end{figure*}

The numerical integration of Eq.~(\ref{eq:LLGS}) was realized with the aid of the projective Euler scheme at a fixed integration step $\delta t = 10$~fs. The computations were performed for the rectangular FL with nanoscale in-plane dimensions $L_1=200$~nm and $L_2 = 80$~nm and the thickness $t_\mathrm{FL} = 1.69$~nm smaller than the threshold thickness $t_\mathrm{SRT} = 1.718$~nm, at which a spin reorientation transition (SRT) to the in-plane magnetization orientation takes place at $V = 0$. The FL demagnetizing factors $N_{ii}$ were calculated analytically~\cite{Aharoni1998} and found to be $N_{11} = 0.0131$, $N_{22} = 0.0336$, and $N_{33} = 0.9533$ (in our case $N_{12} = N_{13} = N_{23} = 0$). Other FL parameters and the conductance of the junction with a typical MgO thickness $t_b = 1$~nm are listed in Table~\ref{tab:param}. Note that the magnetic anisotropy associated with the Co$_{20}$Fe$_{60}$B$_{20}|$overlayer interface~\cite{Peng2017} was neglected in comparison with that of the Co$_{20}$Fe$_{60}$B$_{20}|$MgO one, and the IEC energy was evaluated via the relation $U_\mathrm{IEC} \approx 5.78 \exp{(-7.43 \times 10^9 \mathrm{m}^{-1}t_b)}$~mJ~m$^{-2}$~\cite{Skowronski2010}.

\begin{table}[h]
\caption{\label{tab:param}
Parameters of CoFeB free layer and CoFeB/MgO/CoFeB tunnel junction used in numerical calculations.}
\begin{ruledtabular}
\begin{tabular}{ccc}
Parameter&Value&Reference\\
\hline
$M_s$ & $1.13 \times 10^6$~A~m$^{-1}$&\cite{Lee2011} \\
$\alpha_0$ & 0.01 & \cite{Ikeda2010}\\
$K_1$ & 5~kJ~m$^{-3}$ & \cite{Hall1960}\\
$K_s^0$ & $-1.3 \times 10^{-3}$~J~m$^{-2}$ & \cite{Ikeda2010}\\
$k_s$ & $31$~fJ~V$^{-1}$~m$^{-1}$ & \cite{Alzate2014}\\
$G_\mathrm{P}$ & $1.778 \times 10^{10}$~S~m$^{-2}$ & \cite{Tsunekawa2005}\\
$\eta$ & 0.577 & \cite{Ikeda2010}\\
$p_\mathrm{FL}$ & 0.53 & \cite{Huang2008}\\
$U_\mathrm{IEC}$ & $3.4$~$\mu$J~m$^{-2}$ & \cite{Skowronski2010}
\end{tabular}
\end{ruledtabular}
\end{table}

The magnetization dynamics was first quantified for the Co$_{20}$Fe$_{60}$B$_{20}$ FL covered by the Au layer with the thickness $t_\mathrm{Au}=200$~nm. The voltage drop across the Au overlayer was neglected, because its resistance is much smaller than the MTJ resistance. The displacement current $I_C = C dV_\mathrm{ac}/dt$ proportional to the junction's capacitance $C = \varepsilon_0 \varepsilon_\mathrm{MgO} A / t_b$ ($A$ is the MTJ area, $\varepsilon_\mathrm{MgO} = 9.8$ is the barrier permittivity~\cite{Fontanella1974}) was found to be insignificant in comparison with the tunnel current $I_\mathrm{tun} \geq G_\mathrm{AP} A V_\mathrm{ac}$, because the ratio $2 \pi f C / (G_\mathrm{AP} A)$ is less than 10\% even at the highest studied frequency $\nu=1.7$~GHz. Since gold is a good heat conductor, we also ignored the FL heating caused by the microwave current, which leads to significant heat-driven spin torques in the CoFeB/MgO/FeB/MgO heterostructure~\cite{Goto2019}.

To evaluate the reflection spin-mixing conductance $g^r_{\uparrow \downarrow}$ of the Co$_{20}$Fe$_{60}$B$_{20}|$Au interface, we used the theoretical estimate obtained for the Fe$|$Au one~\cite{Zwierzycki2005}. Taking $\mathrm{Re} \big[g^r_{\uparrow \downarrow}\big] = 1.2 \times 10^{19}$~m$^{-2}$, from Eq.~(\ref{eq:SPrenorm}) we obtained $\alpha = 0.019$ for the renormalized damping parameter. Since $\mathrm{Im} \big[g^r_{\uparrow \downarrow}\big]$ should be negligible at the considered FL thickness $t_\mathrm{FL}=1.69$~nm, which is well above a few-monolayer range, the parameter $\gamma$ was set equal to $\gamma_0$. It should be noted that the spin backflow into FL caused by the spin accumulation in the overlayer reduces the renormalized damping parameter $\alpha$~\cite{JiaoBauer2013}. However, this reduction increases the amplitude of magnetization oscillations and may be ignored in the first approximation at the considered thickness $t_\mathrm{Au} = 200$~nm, which is much larger than the spin diffusion length $\lambda_\mathrm{sd}=35$~nm in Au~\cite{Mosendz2010PRB}.

The numerical calculations were focused on the determination of the frequency dependence of magnetization precession at different amplitudes $V_\mathrm{max}$ of the applied microwave voltage. As a suitable characteristic of the precession magnitude, we employed the sweep $\Delta m_1 = m_1^\mathrm{max} - m_1^\mathrm{min}$ of the scalar product $\mathbf{m} \cdot \mathbf{m}_\mathrm{RL}$ that governs the MTJ conductance $G = G_\mathrm{P} (1 + \eta^2 \mathbf{m} \cdot \mathbf{m}_\mathrm{RL}) / (1 + \eta^2)$. It was found that, at small voltages $V_\mathrm{max} \leq 20$~mV, the dependence $\Delta m_1(f)$ involves a strong symmetric peak situated at the resonance frequency $f_\mathrm{res} \simeq 1.33$~GHz and a finite number of minute peaks located at frequencies $f_n = f_\mathrm{res} / n$ ($n = 2, 3, 4, …$). As the voltage $V_\mathrm{max}$ increases, all peaks grow, gradually become asymmetric and shift to lower frequencies $f_\mathrm{res}(V_\mathrm{max})$ and $f_n(V_\mathrm{max}) \neq f_\mathrm{res}(V_\mathrm{max})/n$ [see Fig.~\ref{fig:res}(a)]. Remarkably, the steep segment of $\Delta m_1(f)$ just below the frequency $f_\mathrm{res}$ of the main peak breaks above a threshold amplitude $V_\mathrm{max} = V_\mathrm{th} \approx 205$~mV [see Fig.~\ref{fig:res}(b)]. When $V_\mathrm{max}$ increases up to about 600~mV, similar breaks appear at the frequencies $f_n$ of the secondary peaks and a second break of the main peak emerges at $f > f_\mathrm{res}$ [Fig.~\ref{fig:res}(c)].

Peculiar dependences of the precession amplitude on the frequency of applied voltage, which appear at $V_\mathrm{max} > V_\mathrm{th}$, are solely due to nonlinearity of the function $\Delta F(\mathbf{m})$ defined by Eq.~(\ref{eq:F}). Indeed, when only the STT and some fixed effective field $\mathbf{H}_\mathrm{eff}$ are taken into account in the numerical calculations, the dependence $\Delta m_1(f)$ assumes the standard form with a single symmetric peak situated at corresponding resonance frequency. Furthermore, the distorted shape of the main peak and the hysteresis shown in Fig.~\ref{fig:res}(b) can be explained by considering the effective field $\mathbf{H}_\mathrm{eff}$ involved in Eq.~(\ref{eq:LLGS}). In the first approximation, the out-of-plane component of $\mathbf{H}_\mathrm{eff}$ can be written as

\begin{equation}
    H_3^\mathrm{eff} = -\bigg(\frac{2 K_s}{\mu_0 M_s t_\mathrm{FL}} + M_s N_{33}\bigg) m_3,
    \label{eq:H3}
\end{equation}
\noindent
where the sum in the brackets is negative owing to the prevailing perpendicular anisotropy created by the CoFeB$|$MgO interface. If the excitation frequency approaches $f_\mathrm{res}$ from below, the precession sweep $\Delta m_1(f)$ increases, which reduces the average direction cosine $\langle m_3 \rangle$ of the precessing magnetization. Hence according to Eq.~(\ref{eq:H3}) the average effective field $\langle H^\mathrm{eff}_3 \rangle$ decreases, leading to a reduction in the resonance frequency $f_\mathrm{res}$ of the large-angle precession. As a result, the precession amplitude increases further and further and eventually jumps to the right branch of the resonance curve at some frequency $f_\mathrm{up}$. On the other hand, when the excitation frequency decreases towards $f_\mathrm{res}$ from above, the accompanying increase of $\Delta m_1(f)$ and reduction of $\langle m_3 \rangle$ lower the resonance frequency. This effect extends the right branch of the resonance curve to frequencies below $f_\mathrm{up}$, but the precession amplitude drops down to the left branch at some frequency  $f_\mathrm{down} < f_\mathrm{up}$, because small decrease of $f_\mathrm{res}$ does not compensate further reduction of $f$. The above considerations explain the hysteresis of $\Delta m_1(f)$ and the position of the break on the frequency scale.

\begin{figure}[b]
\centering
\includegraphics[width=0.85\linewidth]{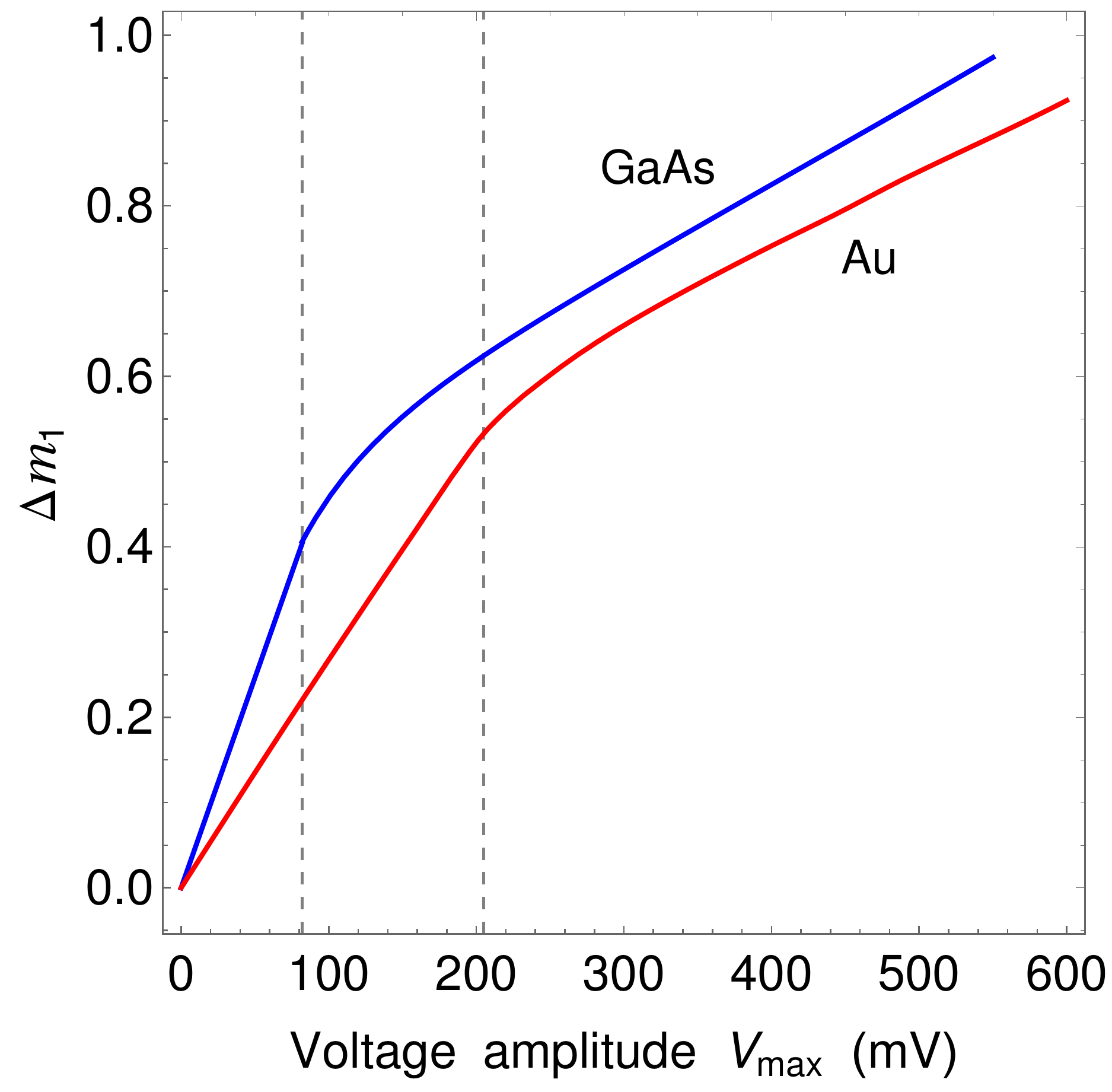}
\caption{\label{fig:dm1(Vmax)} Height $\Delta m_1(f_\mathrm{res})$ of the main peak in the frequency dependence of magnetization precession plotted as a function of the amplitude $V_\mathrm{max}$ of applied microwave voltage. Red and blue curves show results obtained for the free CoFeB layer covered by Au and GaAs, respectively. Vertical dashed lines indicate the threshold voltage $V_\mathrm{th}$. }
\end{figure}

\begin{figure*}[!httb]
\centering
\includegraphics[width=0.95\linewidth]{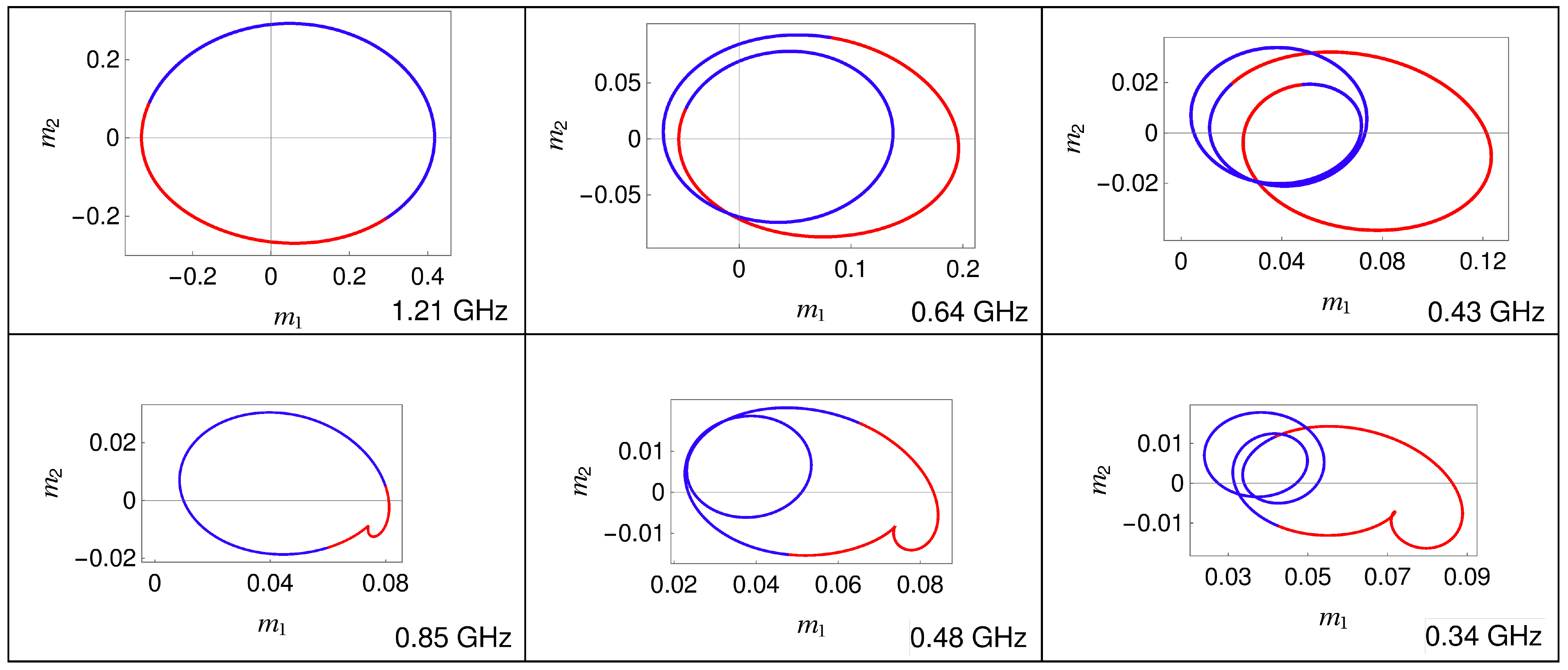}
\caption{\label{fig:trajectories} Trajectories of the end of the unit vector \textbf{m} during the magnetization precession in the free CoFeB layer covered by Au. Curves show projections of these trajectories on the $(x_1, x_2)$ plane parallel to the CoFeB surfaces. Trajectories are calculated at different frequencies $f$ of the applied microwave voltage $V_\mathrm{ac}$ with the amplitude $V_\mathrm{max} = 400$~mV, which are indicated in the figure. The upper row show trajectories arising at the excitation frequencies $f_\mathrm{res}$ and $f_n$ ($n = 2, 3$) corresponding to the peaks of $\Delta m_1(f)$, whereas the lower row presents those forming at frequencies between $f_\mathrm{res}$, $f_2$ and $f_3$. Positive and negative values of the applied voltage $V_\mathrm{ac}(t)$ are indicated by red and blue color, respectively. }
\end{figure*}
The distortion of the main peak, which we revealed for the large-angle magnetization precession in the ultrathin CoFeB layer with perpendicular anisotropy, corresponds to the behavior of a Duffing oscillator with a softening nonlinearity~\cite{Nayfeh1979}. A similar ``foldover'' effect was observed for the STT-driven magnetization dynamics in the Co/Ni multilayer excited by a microwave current~\cite{Chen2009}. However, such multilayer represents a hardening system due to the condition $H^\mathrm{eff}_3 / m_3 < 0$ caused by the prevailing shape anisotropy. Therefore, the resonance curve should have a break on its right branch when plotted on the frequency scale, which is supported by the experimental results~\cite{Chen2009}.

Figure~\ref{fig:dm1(Vmax)} shows the height $\Delta m_1(f_\mathrm{res})$ of the main peak appearing at increasing frequency $f$ as a function of the voltage amplitude $V_\mathrm{max}$. At $V_\mathrm{max} < V_\mathrm{th}$, the peak height varies linearly with $V_\mathrm{max}$, but at higher voltages the variation of $\Delta m_1(f_\mathrm{res})$ becomes nonlinear, which is in line with the significant change of the frequency dependence $\Delta m_1(f)$ appearing above $V_\mathrm{th}$. Remarkably, $\Delta m_1(f_\mathrm{res})$ continues to increase with voltage even near $V_\mathrm{max} = 600$~mV, which is due to strong STT acting on the FL magnetization at the considered small barrier thickness $t_b = 1$~nm. The revealed behavior is different from the voltage dependence of the precession amplitude in the Co$_{20}$Fe$_{60}$B$_{20}$/MgO/Co$_{20}$Fe$_{60}$B$_{20}$ junction with the MgO thickness of 2~nm~\cite{Rana2017}, which saturates at $V_\mathrm{max} \sim 300$~mV due to negligible STT at such $t_b$. It should be noted that, in the range of small voltages $V_\mathrm{max} << V_\mathrm{th}$, the results of our numerical calculations agree with those obtained analytically by solving the linearized LLGS equation. This agreement confirms the validity of our computations. 

To clarify the origin of the secondary peaks, we determined the trajectories of the end of the unit vector \textbf{m} during the magnetization oscillations generated at different excitation frequencies for a representative voltage amplitude $V_\mathrm{max} = 400$~mV. Figure~\ref{fig:trajectories} shows projections of these trajectories on the $(x_1, x_2)$ plane, which arise at the frequencies $f_\mathrm{res}$ and $f_n$ $(n = 2, 3)$ corresponding to the peaks of $\Delta m_1(f)$, in comparison with those formed at frequencies between $f_\mathrm{res}$, $f_2$, and $f_3$. It can be seen that, at $f = f_n$, the magnetization makes $n$ full turns around the equilibrium direction during one period $1/f_n$ of the voltage oscillation. Since $f_n \approx f_\mathrm{res}/n$, the mean period of the forced magnetization precession appears to be close to the period $1/f_\mathrm{res}$ of free oscillations, which explains enhancement of the precession amplitude at frequencies $f_n$. Evidently, the above condition cannot be fulfilled at excitation frequencies significantly differing from $f_\mathrm{res}/n$. It should be noted that the predicted secondary peaks are an attribute of parametric resonance~\cite{Jia2016}, which occurs when the natural oscillation frequency is varied by an external stimulus. Owing to VCMA, the application of microwave voltage modifies the absolute value $|\mathbf{H}_\mathrm{eff}|$ of the effective field, which governs the natural precession frequency. Therefore, the presence of VCMA is responsible for the revealed secondary peaks.

Similar frequency and voltage dependences have been obtained for the electrically driven magnetization precession in the Co$_{20}$Fe$_{60}$B$_{20}$ free layer covered by GaAs. Since the reflection spin-mixing conductance $g^r_{\uparrow \downarrow}$ of the CoFeB$|$GaAs interface is expected to be relatively small in comparison with that of the CoFeB$|$Au one~\cite{Ando2011NM}, the influence of the spin pumping on the parameters $\gamma$ and $\alpha$ involved in Eq.~(\ref{eq:LLGS}) can be ignored. Therefore, the magnetic damping becomes smaller ($\alpha = 0.01$), which leads to higher peaks of $\Delta m_1(f)$ in the free layer covered by GaAs. The voltage dependence of the height $\Delta m_1(f_\mathrm{res})$ of the main peak appearing at increasing the excitation frequency is shown in Fig.~\ref{fig:dm1(Vmax)}.  It is qualitatively similar to the voltage dependence obtained for the Co$_{20}$Fe$_{60}$B$_{20}$ free layer covered by Au, but differs by larger values of $\Delta m_1(f_\mathrm{res})$ and a smaller threshold voltage $V_\mathrm{th} \approx 82$~mV.

\section{\label{sec:into:Au}SPIN INJECTION AND PUMPING\\ INTO METALLIC OVERLAYER}

Using the results obtained for the magnetization dynamics induced by the microwave voltage applied to the Co$_{20}$Fe$_{60}$B$_{20}$/MgO/Co$_{20}$Fe$_{60}$B$_{20}$ junction, we calculated the spin current generated in the Au overlayer near the interface with FL. In our case, such current is the sum of two contributions, which result from the spin pumping caused by the magnetization precession and the spin injection proportional to the spin polarization of the charge current. 
\begin{figure}[b]
\centering
\begin{minipage}[h]{0.9\linewidth}
\includegraphics[width=1\linewidth]{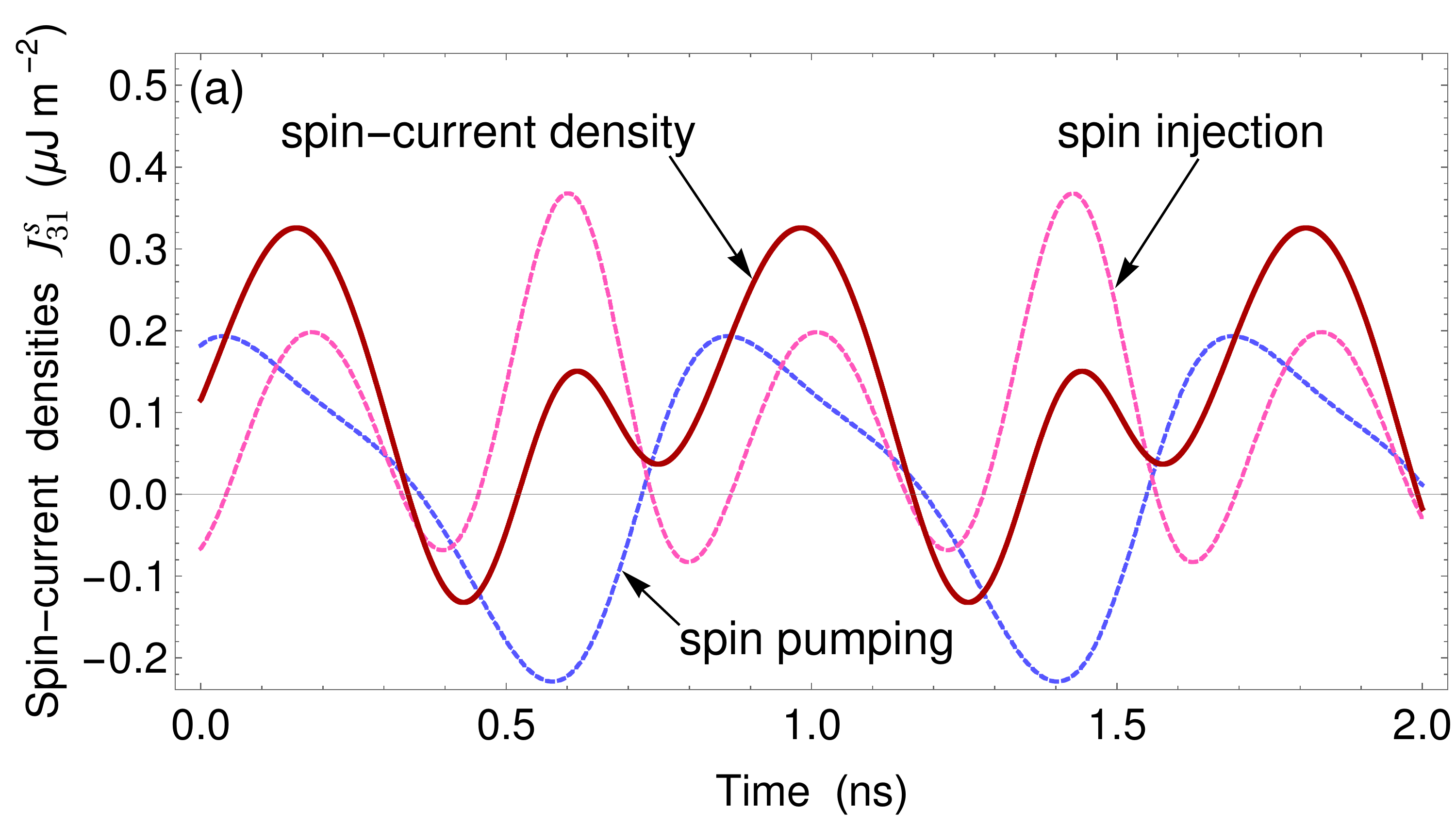}
\end{minipage}
\vfil
\begin{minipage}[h]{0.9\linewidth}
\includegraphics[width=1\linewidth]{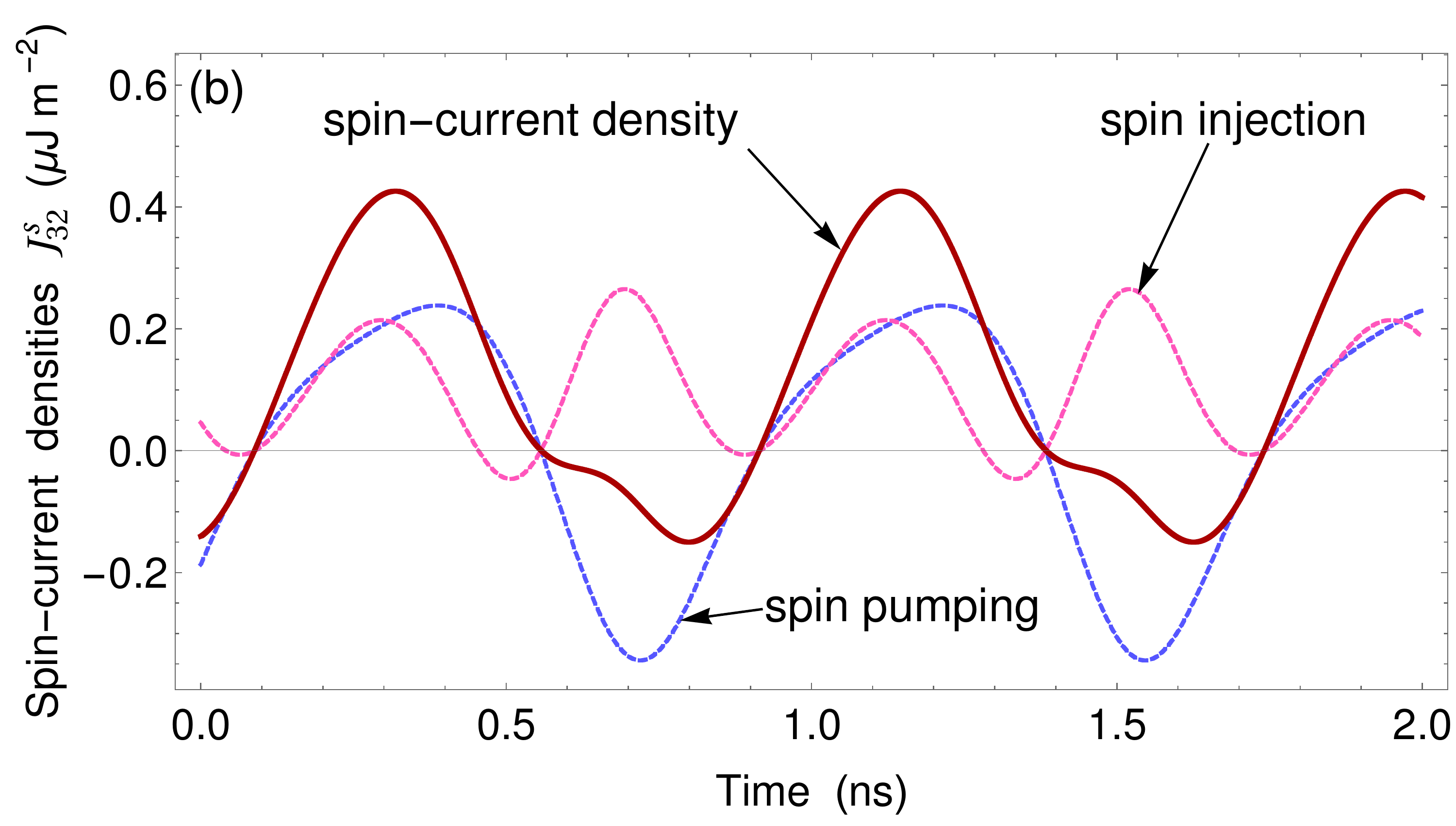}
\end{minipage}
\vfil
\begin{minipage}[h]{1.0\linewidth}
\includegraphics[width=1.0\linewidth]{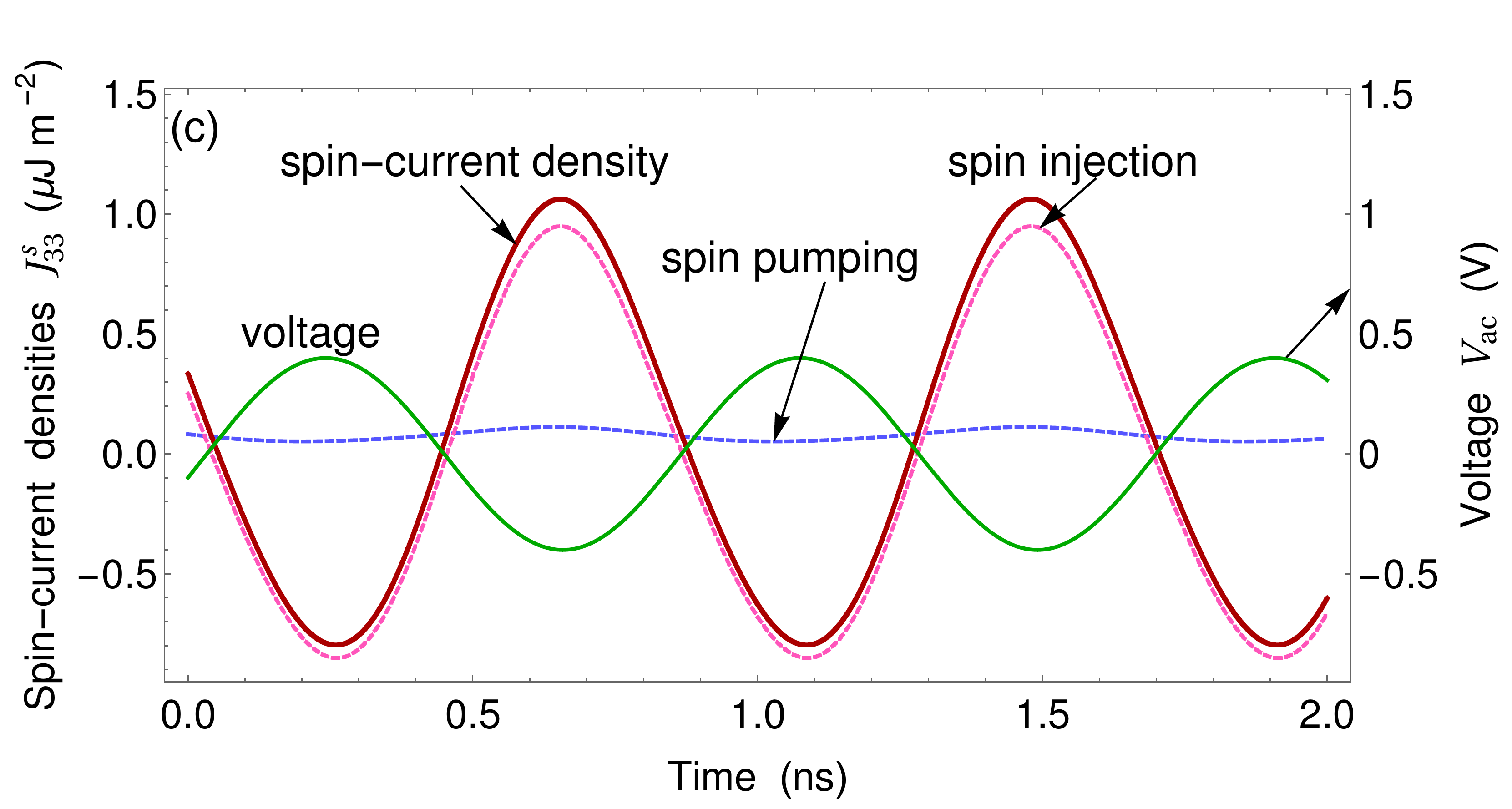}
\end{minipage}
\caption{\label{fig:SC(t)} Time dependences of the total spin-current densities $J^s_{31}$ (a), $J^s_{32}$ (b), and $J^s_{33}$ (c) generated at the Au/CoFeB interface by the microwave voltage with the frequency $f_\mathrm{res} = 1.2$~GHz and amplitude $V_\mathrm{max} = 400$~mV. Contributions of spin-polarized charge current and precession-induced spin pumping are shown by pink and blue lines, respectively.}
\end{figure} 
Since the spin current is characterized by the direction of spin flow and the orientation of spin polarization, we employed a second-rank tensor $\mathbf{J}_s$ to define the spin-current density~\cite{Dyakonov1971}. At the considered FL thickness, the pumped spin-current density $\mathbf{J}_\mathrm{sp}$ near the Co$_{20}$Fe$_{60}$B$_{20}|$Au interface can be evaluated using the formula $\mathbf{e}_n \cdot \mathbf{J}_\mathrm{sp} \simeq (\hbar / 4 \pi) \mathrm{Re} \big[g^r_{\uparrow \downarrow} \big] \mathbf{m} \times d\mathbf{m} / dt$, where $\mathbf{e}_n$ is the unit normal vector to the interface pointing into Au, and $\mathrm{Re} \big[g^r_{\uparrow \downarrow} \big]$ may be set equal to $1.2 \times 10^{19}$~m$^{-2}$~\cite{Zwierzycki2005}. The spin injection created by the charge current with the density $J_c = G V_\mathrm{ac}$ is described by the relation

\begin{equation}
    \mathbf{e}_n \cdot \mathbf{J}_\mathrm{si} \simeq -\frac{\hbar}{2 e} G_\mathrm{P} \frac{1 + \eta^2 \mathbf{m} \cdot \mathbf{m}_\mathrm{RL}}{1 + \eta^2} V_\mathrm{ac} p_\mathrm{FL} \mathbf{m},
    \label{eq:si}
\end{equation}
\noindent
where $p_\mathrm{FL} = (N_\uparrow - N_\downarrow) / (N_\uparrow + N_\downarrow)$ is the spin polarization in the FL having the densities of states $N_\uparrow$ and $N_\downarrow$ of spin-up and spin-down electrons at the Fermi level, respectively. It should be noted that the spin flow along the Co$_{20}$Fe$_{60}$B$_{20}|$Au interface, which is caused by the spin Hall effect, has a negligible magnitude in comparison with that of the spin flow in the direction orthogonal to the interface.

Representative results obtained for variations of the total spin-current density $\mathbf{J} = \mathbf{J}_\mathrm{sp} + \mathbf{J}_\mathrm{si}$ with time are presented in Fig.~\ref{fig:SC(t)}. Since the discussed spin current flows along the $x_3$ axis of our coordinate system shown in Fig.~\ref{fig:mtj}, the only nonzero elements of the tensor $\mathbf{J}_s$ are $J^s_{3k}$ ($k = 1, 2, 3$). Figure~\ref{fig:SC(t)} shows that, at the considered high voltage amplitude $V_\mathrm{max} = 400$~mV, $J^s_{31}(t)$ and $J^s_{32}(t)$ exhibit strongly non-sinusoidal time dependences, whereas $J^s_{33}(t)$ is distinguished by an approximately sinusoidal one. Concerning the contributions of precession-induced spin pumping and spin-polarized charge current, we see that the sweep of $J_{31}^\mathrm{sp}(t)$ is almost the same as the $J_{31}^\mathrm{si}(t)$ one, while the sweep of $J_{32}^\mathrm{sp}(t)$ is noticeably bigger than that of $J_{32}^\mathrm{si}(t)$. In contrast, the sweep of $J_{33}^\mathrm{si}(t)$ is much larger than that of the density $J_{33}^\mathrm{sp}(t)$, which does not change sign and varies with time only slightly (Fig.~\ref{fig:SC(t)}).

\begin{figure}[t]
\centering
\includegraphics[width=0.9\linewidth]{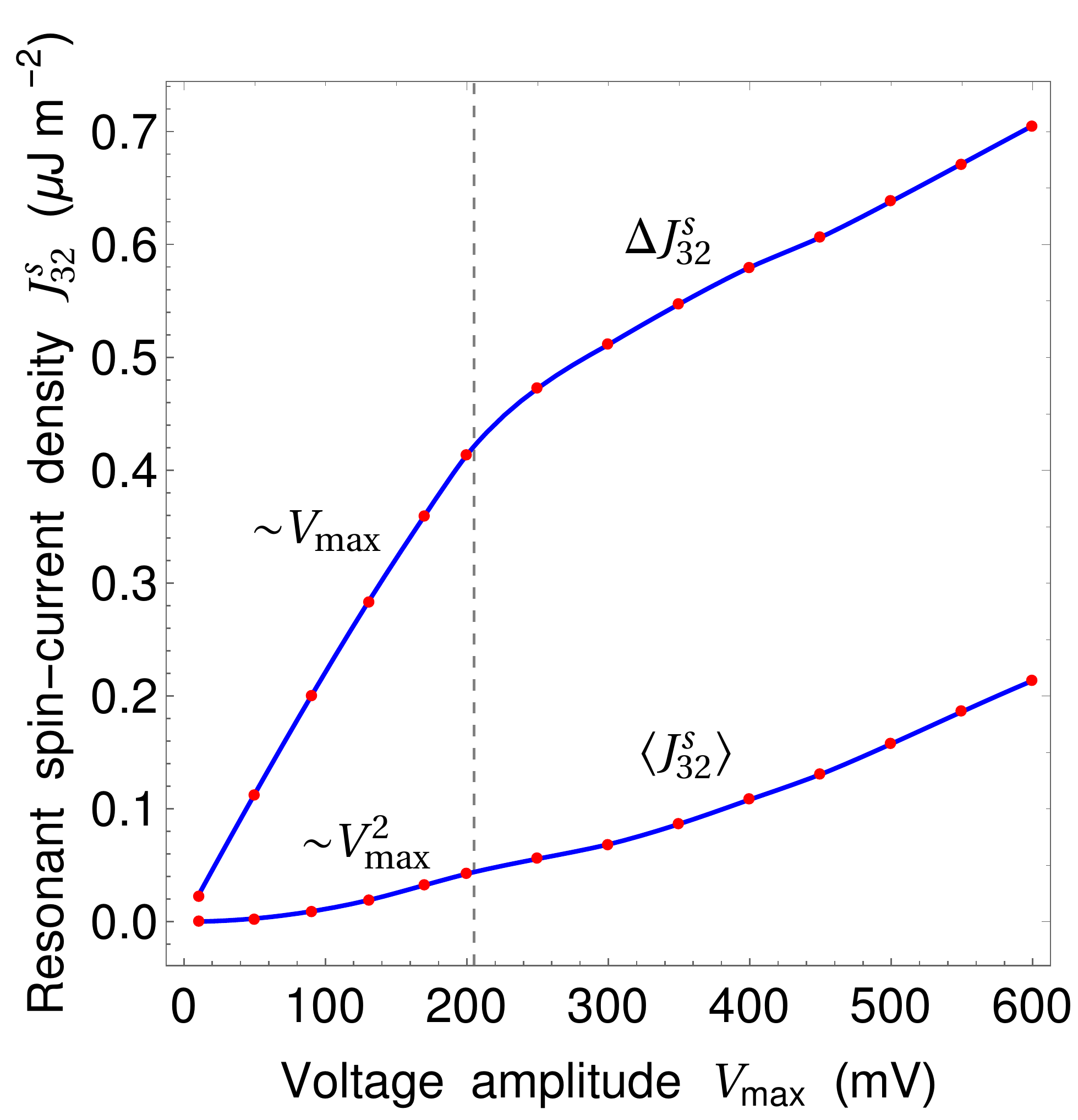}
\caption{\label{fig:Js32(Vmax)} Dependences of the ac and dc components $\Delta J^s_{32}$ and $\langle \Delta J^s_{32} \rangle$ of the spin-current density $J^s_{32}$ on the voltage amplitude $V_\mathrm{max}$, which were evaluated at the voltage-dependent resonance frequency $f_\mathrm{res}(V_\mathrm{max})$. Vertical dashed line indicates the threshold voltage amplitude $V_\mathrm{th}$. }
\end{figure}

The sweep $\Delta J_{3k}^s$ characterizes the ac component of the spin-current density $J_{3k}^s(t)$, and the dc component $\langle J^s_{3k} \rangle$ can be determined by averaging $J_{3k}^s(t)$ over the period $1/f$ of voltage oscillations. Note that $J_{31}^\mathrm{si}(t)$ and $J_{32}^\mathrm{si}(t)$ oscillate with the double excitation frequency $2f$, because the  magnetization projections $m_1(t)$ and $m_2(t)$ involved in Eq.~(\ref{eq:si}) undergo significant variations with time (see Fig.~\ref{fig:trajectories}). Interestingly, the numerical calculations show that the spin injection $J^\mathrm{si}_{32}(t)$ does not significantly influence the ac component of the total spin-current density $J^s_{32}(t)$. Therefore, the ac component of $J^s_{32}(t)$ practically equals that of $J^\mathrm{sp}_{32}(t)$, thus characterising the precession-induced spin current. Furthermore, the dc component of $J^\mathrm{sp}_{32}(t)$ is almost zero so that the time-averaged value of $J^s_{32}(t)$ is governed by the spin injection.
\begin{figure*}[!httb]
\centering
\begin{minipage}[h]{0.29\linewidth}
\includegraphics[width=1\linewidth]{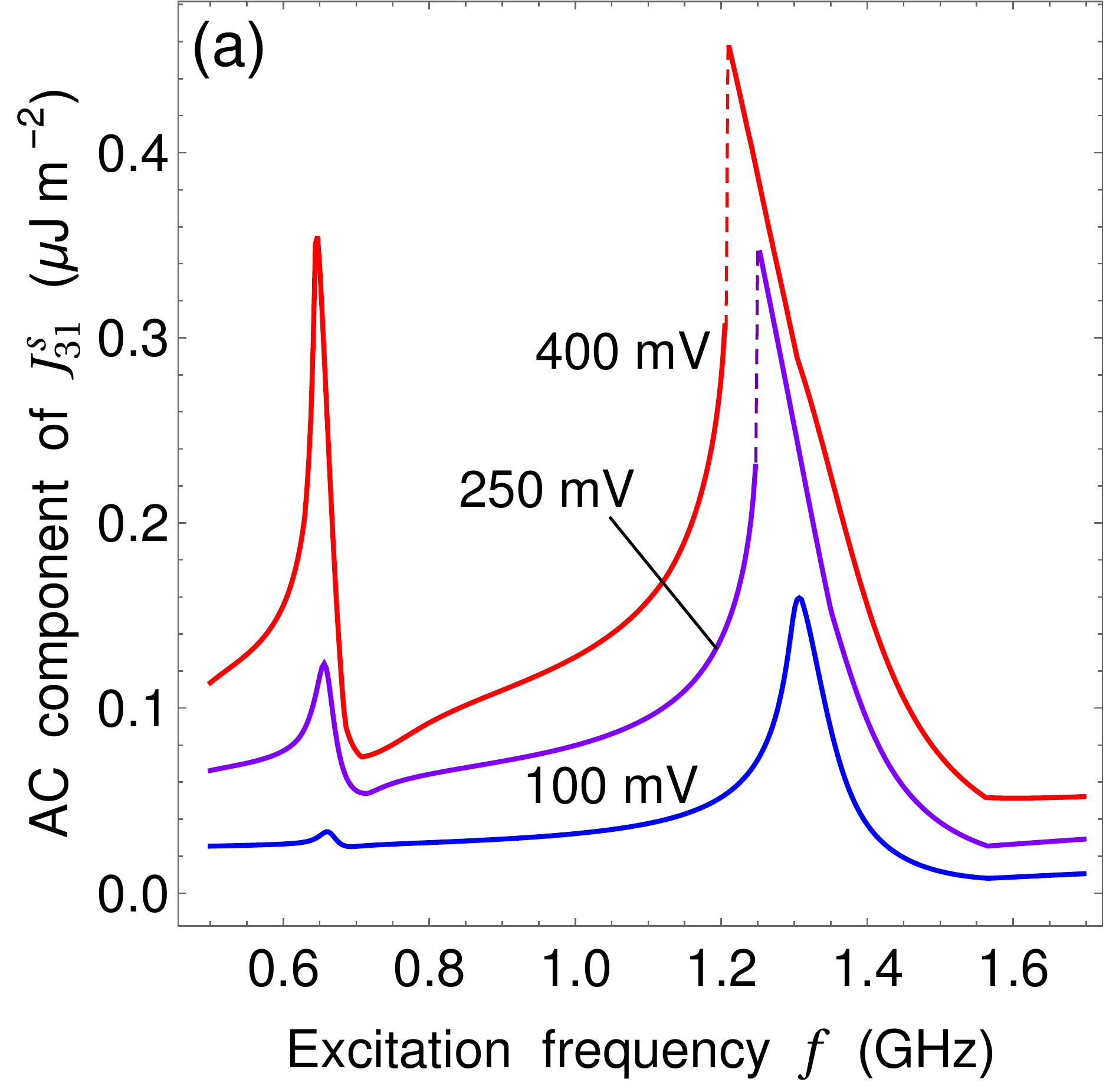}
\end{minipage}
\hfil
\begin{minipage}[h]{0.29\linewidth}
\includegraphics[width=1\linewidth]{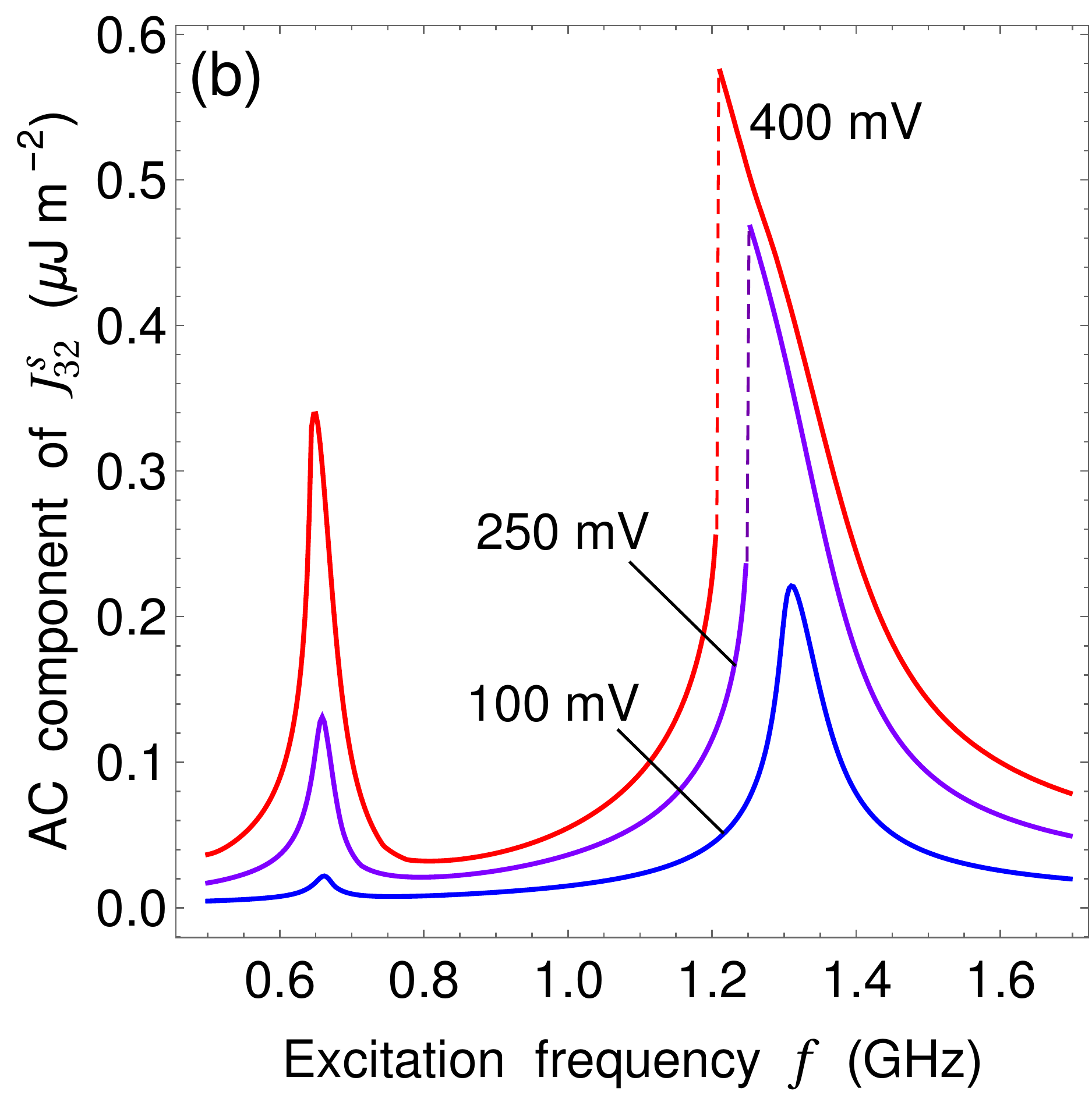}
\end{minipage}
\hfil
\begin{minipage}[h]{0.29\linewidth}
\includegraphics[width=1\linewidth]{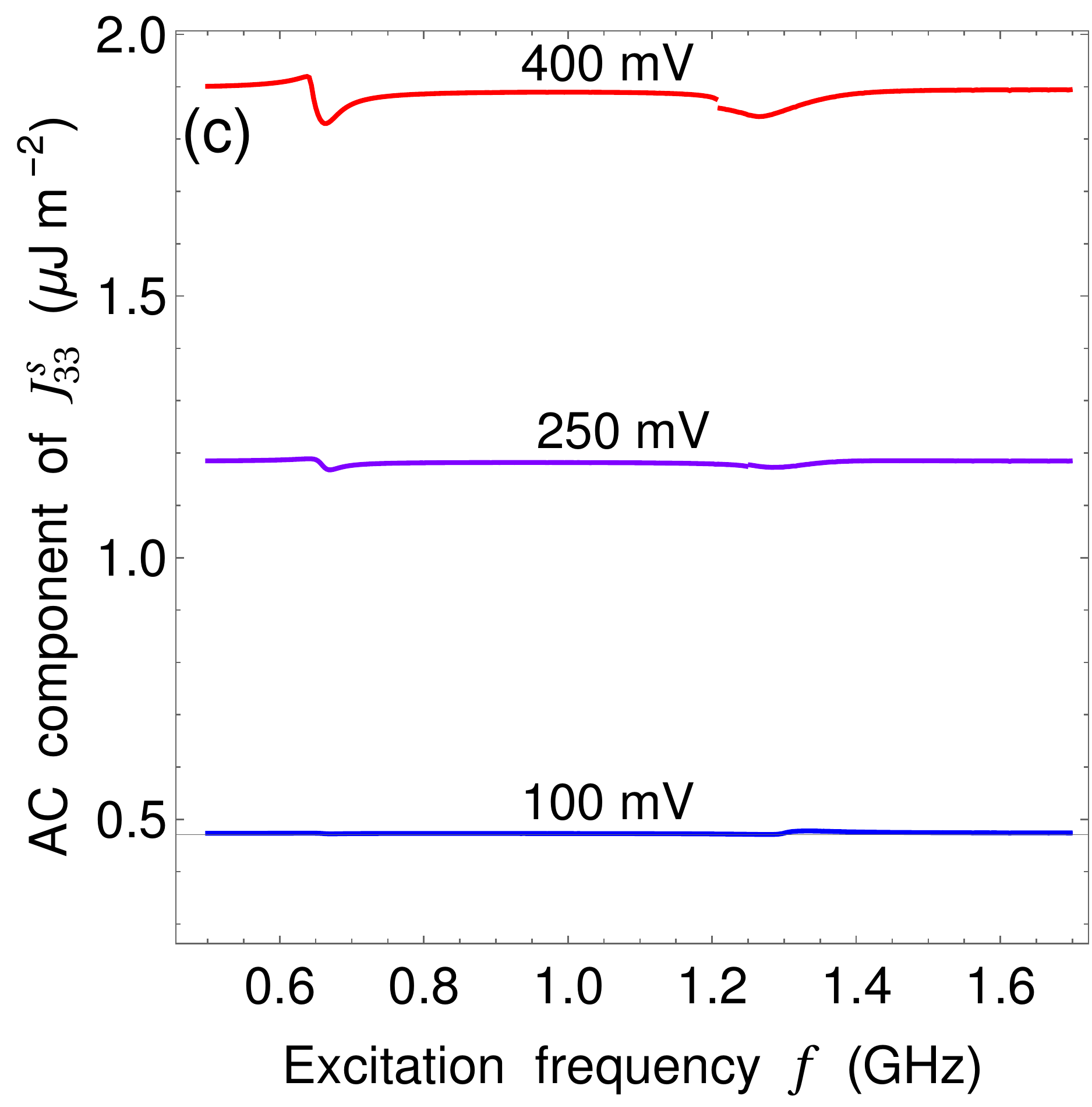}
\end{minipage}
\caption{\label{fig:AC(f)} Frequency dependences of the ac components of spin-current densities $J^s_{31}$ (a), $J^s_{32}$ (b), and $J^s_{33}$ (c) generated at the Au$|$CoFeB interface by microwave voltages with different amplitudes $V_\mathrm{max}$ indicated near the graphs. }
\end{figure*} 
\begin{figure*}[!httb]
\centering
\begin{minipage}[h]{0.29\linewidth}
\includegraphics[width=1\linewidth]{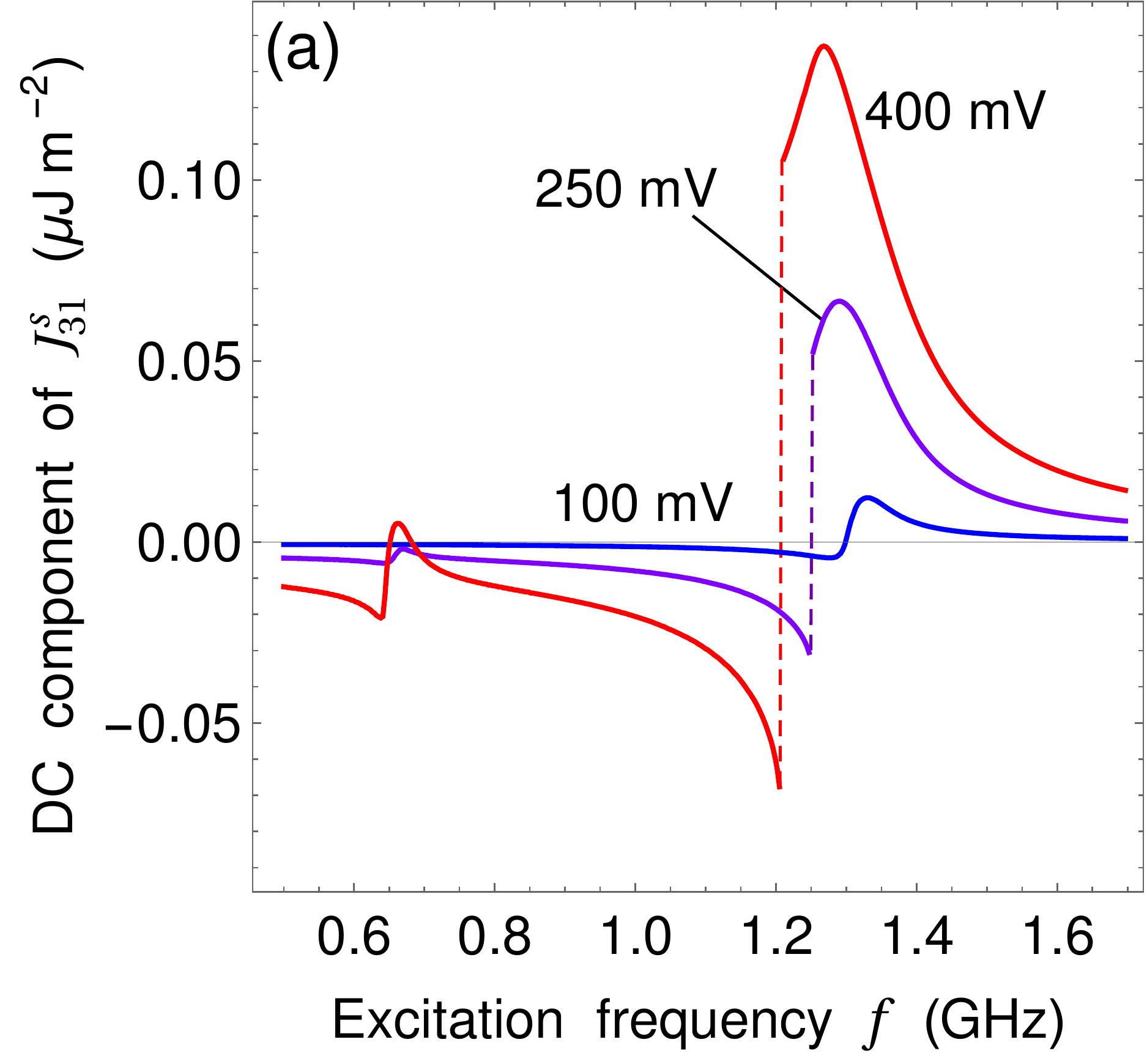}
\end{minipage}
\hfil
\begin{minipage}[h]{0.29\linewidth}
\includegraphics[width=1\linewidth]{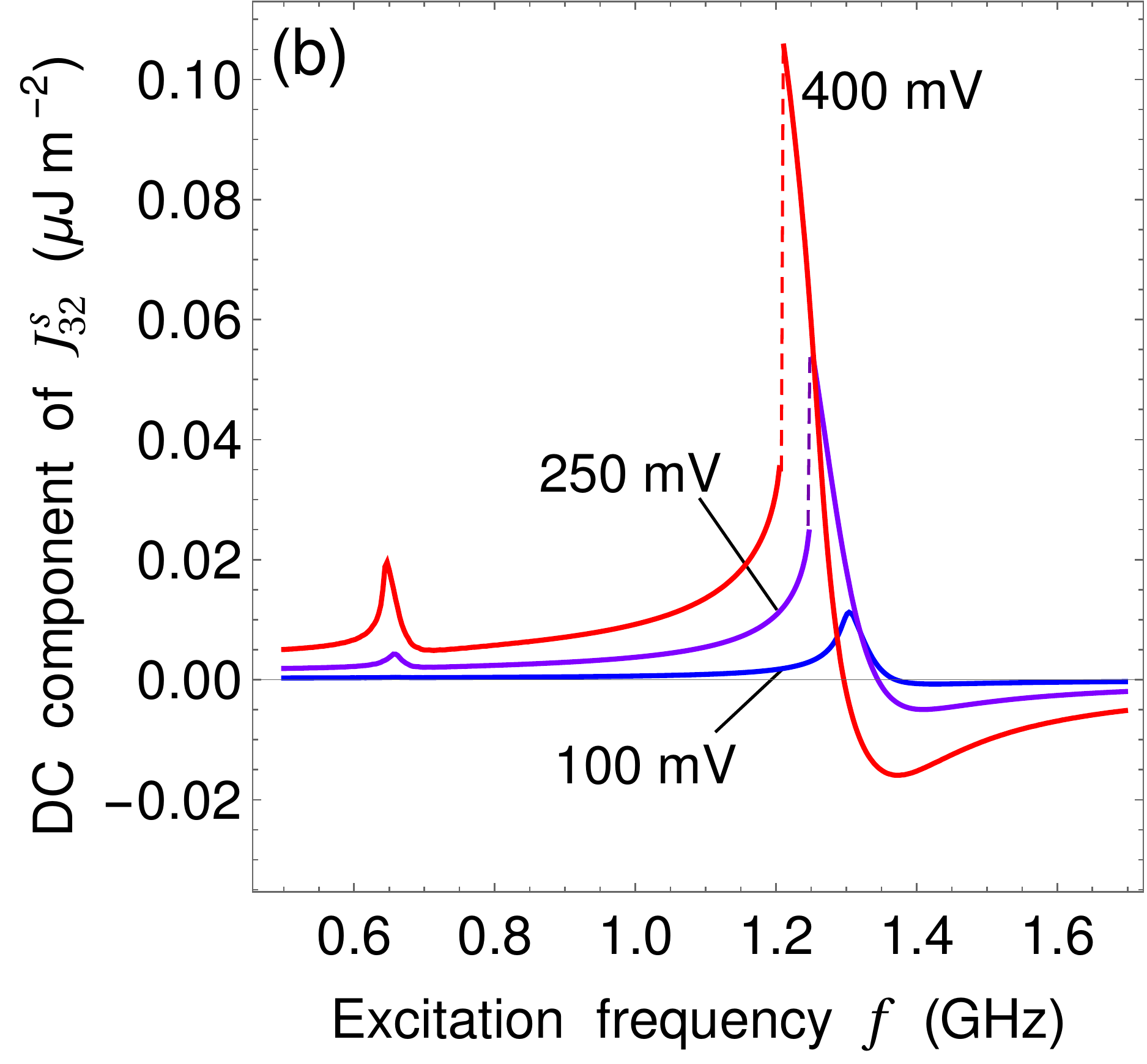}
\end{minipage}
\hfil
\begin{minipage}[h]{0.29\linewidth}
\includegraphics[width=1\linewidth]{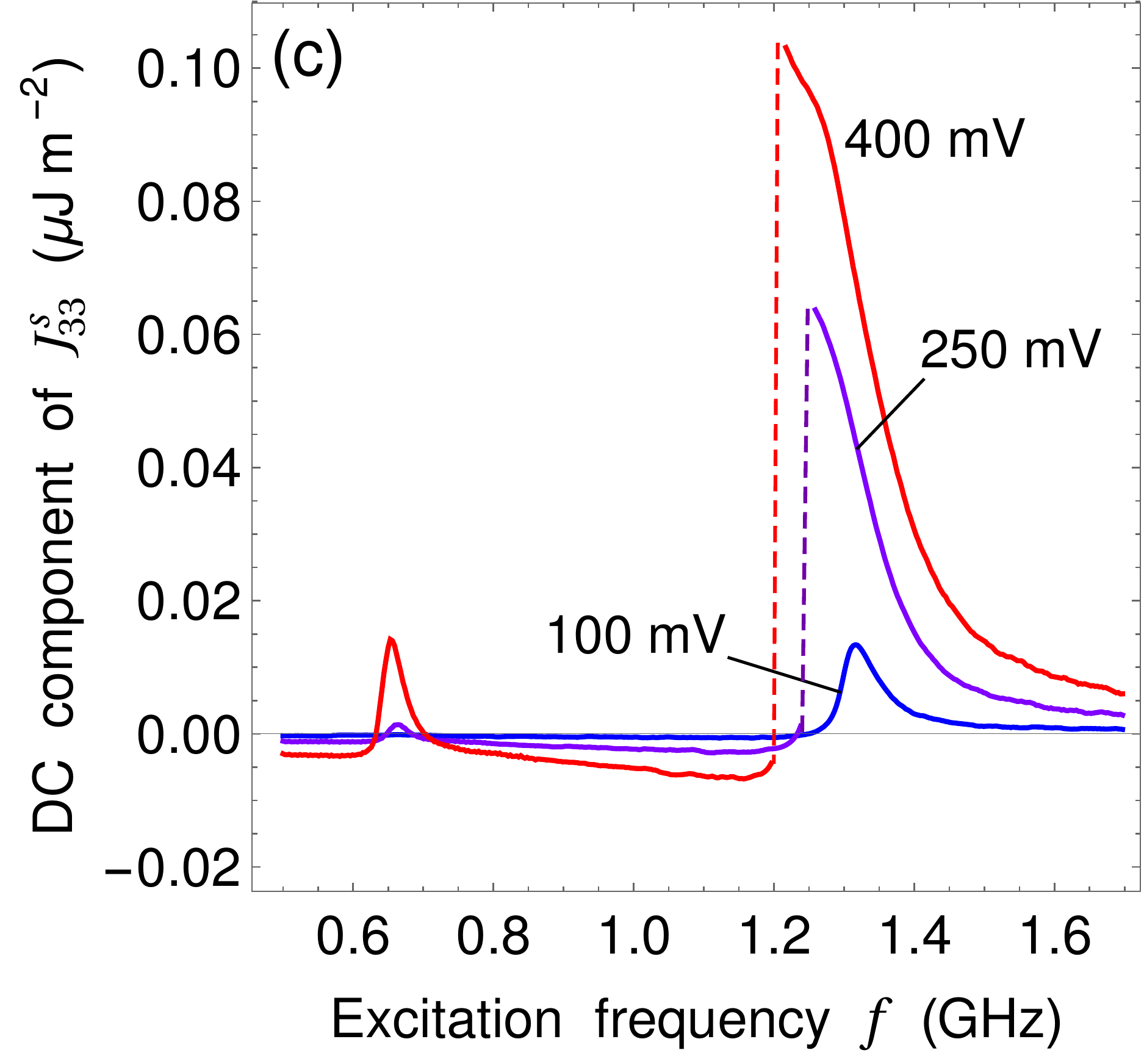}
\end{minipage}
\caption{\label{fig:DC(f)} Frequency dependences of the dc components of spin-current densities $J^s_{31}$ (a), $J^s_{32}$ (b), and $J^s_{33}$ (c) generated at the Au$|$CoFeB interface by microwave voltages with different amplitudes $V_\mathrm{max}$ indicated near the graphs. }
\end{figure*} 
Figure~\ref{fig:AC(f)} shows variations of the ac components $\Delta J_{31}^s$, $\Delta J_{32}^s$, and $\Delta J_{33}^s$ with the excitation frequency $f$, which take place near the resonance frequency $f_\mathrm{res}$ of the magnetization precession. It can be seen that $\Delta J_{31}^s$ and $\Delta J_{32}^s$ have a strong peak at $f_\mathrm{res}$, while $\Delta J_{33}^s$ only weakly depends on the excitation frequency. Remarkably, the secondary peak of $\Delta J_{31}^s$ and $\Delta J_{32}^s$, which is located at $f \approx f_\mathrm{res} / 2$, rapidly grows with increasing voltage amplitude $V_\mathrm{max}$ and becomes comparable in magnitude with the main peak at $V_\mathrm{max} \geq 400$~mV. In contrast, the dc spin-current densities $\langle J_{3k}^s \rangle$ exhibit strong peaks near $f_\mathrm{res}$ only (see Fig.~\ref{fig:DC(f)}). To illustrate the influence of the voltage amplitude $V_\mathrm{max}$ on the spin-current density generated near the CoFeB$|$Au interface, we presented in Fig.~\ref{fig:Js32(Vmax)} the dependences $\Delta J_{32}^s(V_\mathrm{max})$ and $\langle J_{32}^s \rangle(V_\mathrm{max})$ evaluated at the voltage-dependent resonance frequency $f_\mathrm{res}(V_\mathrm{max})$. Below the threshold amplitude $V_\mathrm{th} \approx 205$~mV, the ac component varies with $V_\mathrm{max}$ almost linearly, whereas the dc component follows the power law $\langle J^s_{32} \rangle \propto V^2_\mathrm{max}$. At higher voltage amplitudes, both $\Delta J_{32}^s$ and $\langle J_{32}^s \rangle$ increase slower than expected from the above dependences, but at $V_\mathrm{max} = 600$~mV they reach significantly enhanced values of about 0.7 and 0.2~$\mu$J~m$^{-2}$, respectively. It should be noted that these values differ strongly from the spin-current densities $\Delta J_{32}^s \sim 0.1$~$\mu$J~m$^{-2}$ and $\langle J^s_{32} \rangle \sim 2$~$\mu$J~m$^{-2}$ generated by the Co$_{20}$Fe$_{60}$B$_{20}$/MgO/Co$_{20}$Fe$_{60}$B$_{20}$ MTJ excited by direct charge current~\cite{PRB2019}.

The spin-current densities $J^s_{3k}$ presented in Figs.~\ref{fig:SC(t)}-\ref{fig:Js32(Vmax)} are partially suppressed at the interface by the spin backflow $\mathbf{J}_\mathrm{sb}.$ The product $\mathbf{e}_n \cdot \mathbf{J}_\mathrm{sb}$ consists of the components parallel and perpendicular to the magnetization direction \textbf{m}, which are caused by spin injection and spin pumping, respectively [see Eq.~(\ref{eq:si}) and expression for spin pumping]. Numerical estimates show that the conductance mismatch problem does not appear for the CoFeB$|$Au interface. Accordingly, the longitudinal spin backflow is negligible compared to $\mathbf{J}_\mathrm{si}$ in our case. In contrast, the transverse spin backflow reduces the spin pumping significantly. To calculate the actual spin-current density $\mathbf{J}_\mathrm{Au}(x_3)$ in Au, we solve the spin diffusion equation with appropriate boundary conditions for spin injection and pumping. The boundary condition at $x_3=t_\mathrm{Au}$ represents zero spin flux. The boundary condition at the interface reads $\mathbf{e}_n \cdot \mathbf{J}_\mathrm{Au} = \mathbf{e}_n \cdot \mathbf{J}^\mathrm{si} + \mathbf{e}_n \cdot \mathbf{J}^\mathrm{sp} - \mathrm{Re}\big[g^r_{\uparrow \downarrow}\big]\bm{\mu}_s/4\pi$~\cite{Tserkovnyak2005}, where $\bm{\mu}_s$ is the spin accumulation in Au near the interface. It should be noted that in our case $\bm{\mu}_s = 2 k_B T (N_\uparrow - N_\downarrow) / (N_\uparrow + N_\downarrow)$, where $N_\uparrow$ and $N_\downarrow$ are the densities of states in Au, which characterize spin-up and spin-down electrons at the Fermi level. The introduction of spin accumulation allows to express the diffusive spin-current density in Au as $\mathbf{J}_\mathrm{Au} = -[\sigma_\mathrm{Au} \hbar/(4 e^2)]\partial \bm{\mu}_s / \partial \mathbf{r}$~\cite{Tserkovnyak2002PRB}, where $\sigma_\mathrm{Au}$ is the electrical conductivity of Au. Solving the diffusion equation in the adiabatic approximation~\cite{Tserkovnyak2002PRB}, we obtain an analytic relation

\begin{equation}
\begin{gathered}
    \mathbf{J}_\mathrm{Au}(x_3) = \frac{\sinh{[(t_\mathrm{Au}-x_3) / \lambda_\mathrm{sd}]}}{\sinh{[t_\mathrm{Au} / \lambda_\mathrm{sd}]}}\\ \bigg[\mathbf{J}_\mathrm{si} + \mathbf{J}_\mathrm{sp} \Big( 1 + \mathrm{Re}\big[g^r_{\uparrow \downarrow}\big] \displaystyle \frac{\lambda_\mathrm{sd} e^2}{\pi \sigma_\mathrm{Au} \hbar} \coth{\frac{t_\mathrm{Au}}{\lambda_\mathrm{sd}}}  \Big)^{-1} \bigg],
\end{gathered}
\label{eq:sc(x3)}
\end{equation}
\noindent
which describes how  $\mathbf{J}_\mathrm{Au}$ decays with the distance $x_3$ from the CoFeB$|$Au interface due to spin relaxation and diffusion. Equation~(\ref{eq:sc(x3)}) is similar to the formula presented in~\cite{Mosendz2010PRB}, but differs by the spin-current density at the interface $\mathbf{J}_\mathrm{si} + \mathbf{J}_\mathrm{sp} \beta$, where $\beta$ is the backflow factor. Taking $\sigma_\mathrm{Au} = 4.5 \times 10^7$~S~m$^{-1}$~\cite{Haynes2009}, we find $\beta \approx 0.6$ so that the actual spin pumping into Au is smaller than $\mathbf{J}_\mathrm{sp}$ by about 40\%.

Our theoretical results demonstrate that the CoFeB/MgO/CoFeB tunnel junction subjected to a microwave voltage with an appropriate frequency represents a promising spin injector into normal metals. Since the generated spin current creates a charge current owing to the inverse spin Hall effect (ISHE), the efficiency of spin injection can be probed electrically~\cite{Saitoh2006}. Motivated by this opportunity, we calculated distributions of the charge-current density $\mathbf{J}_c(\mathbf{r}, t)$ and electric potential $\phi(\mathbf{r},t)$ in the Au overlayer in the quasistatic approximation~\cite{PRB2019}. As the ISHE contribution $\mathbf{J}_\mathrm{ISHE}$ to the density $\mathbf{J}_c$ is governed by the vector product of the spin accumulation $\bm{\mu}_s$ and the unit vector $\mathbf{e}_s$ directed along the spin flow~\cite{Mosendz2010PRL}, the element $J^\mathrm{Au}_{33}$ of the spin-current-density tensor $\mathbf{J}_\mathrm{Au}$ does not affect $\mathbf{J}_\mathrm{ISHE}$ and can be disregarded. Further, the elements $J^\mathrm{Au}_{31}$ and $J^\mathrm{Au}_{32}$ create contributions only to the projections of $\mathbf{J}_\mathrm{ISHE}$ on the orthogonal axes $x_2$ and $x_1$, respectively. Restricting our calculations to the determination of the charge transport in the $(x_1, x_3)$ plane, which governs the transverse voltage $V_1(x_3) = \phi(x_1 = L_1, x_3) - \phi(x_1 = 0, x_3)$ between the sides of Au overlayer normal to the $x_1$ axis (Fig.~\ref{fig:mtj}), we may disregard $J^\mathrm{Au}_{31}$ as well. For the relevant projection $J_1^\mathrm{ISHE}$ of the ISHE current density, the theory gives $J_1^\mathrm{ISHE}(x_1) = \alpha_\mathrm{SH}(2e / \hbar)J^\mathrm{Au}_{32}(x_3)$, where $\alpha_\mathrm{SH}=0.0035$ is the spin Hall angle of Au~\cite{Mosendz2010PRB}, and the spin-current density $J^\mathrm{Au}_{32}$ is determined by Eq.~(\ref{eq:sc(x3)}). 

The total density $\mathbf{J}_c$ of the charge current flowing in the Au layer is the sum of the ISHE contribution $\mathbf{J}_\mathrm{ISHE}$ and the drift contribution $\mathbf{J}_\mathrm{drift} = -\sigma_\mathrm{Au} \nabla \phi$. To calculate the electric potential $\phi(\mathbf{r}, t)$, we used the Laplace's equation $\nabla^2 \phi = 0$ appended by appropriate boundary conditions. Namely, the charge-current density $J_3^c$ at the CoFeB$|$Au interface $x_3 = 0$ was set equal to the density $J_c = G V_\mathrm{ac}$ of the tunnel current, which was assumed uniform, because the anomalous Hall effect in the CoFeB layer has a weak effect on the transverse voltage $V_1(x_3)$~\cite{PRB2019}.
\begin{figure}[t]
\centering
\includegraphics[width=0.9\linewidth]{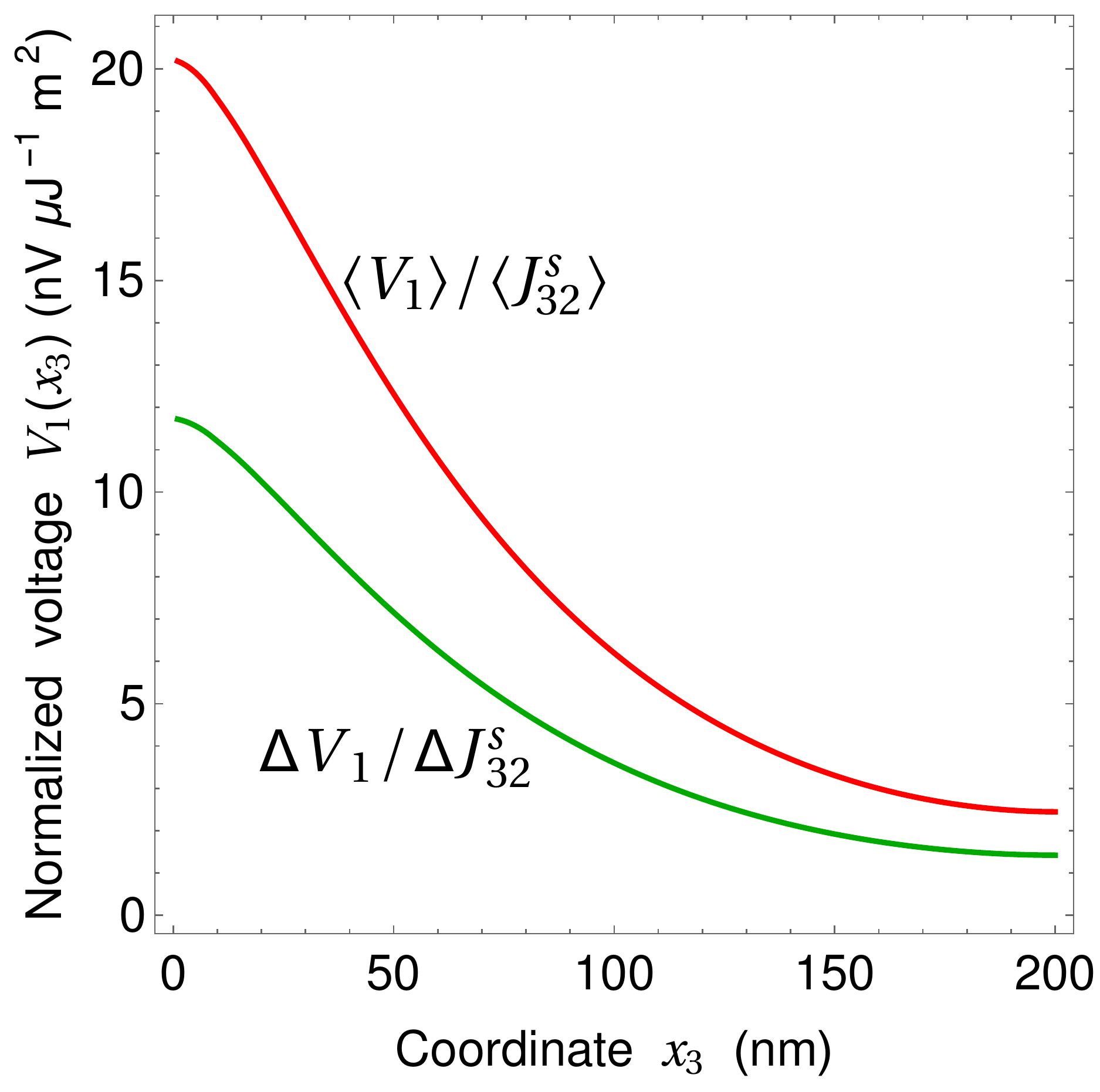}
\caption{\label{fig:Vtr(x3)} Normalized transverse voltage between the sides of Au overlayer normal to the $x_1$ axis plotted as a function of the distance $x_3$ from the Au$|$CoFeB interface.}
\end{figure}
The same boundary condition was introduced at the upper boundary $x_3 = t_\mathrm{Au}$ of the Au overlayer to exclude the charge accumulation in Au. At the side boundaries $x_1 = 0$ and $x_1 = L_1$ of the overlayer, the total current density $J_3^c$ should go to zero, which yields $\sigma_\mathrm{Au}\partial \phi / \partial x_1 = J_1^\mathrm{ISHE}$, where $J_1^\mathrm{ISHE}$ is directly proportional to the spin-current density  given by Eq.~(\ref{eq:sc(x3)}). Remarkably, because of this condition the transverse voltage $V_1(x_3, t)$ at any time moment is a linear combination of $J_{32}^\mathrm{sp}(t)$ and $J_{32}^\mathrm{si}(t)$ presented in Fig.~\ref{fig:SC(t)}. Taking into account that $\Delta (J^\mathrm{si}_{32}+J^\mathrm{sp}_{32}) \approx \Delta J^\mathrm{sp}_{32}$ and $\langle J^\mathrm{si}_{32}+J^\mathrm{sp}_{32}\rangle \approx \langle J^\mathrm{si}_{32} \rangle$, we find that the ac component $\Delta V_1(x_3)$ of the transverse voltage is proportional to spin pumping, whereas the dc component $\langle V_1 \rangle(x_3)$ is proportional to spin injection. Hence a certain tunneling heterostructure has the universal dependences of the normalized voltages $\Delta V_1(x_3) / \Delta J_{32}^s$ and $\langle V_1 \rangle(x_3) / \langle J_{32}^s \rangle$ on the distance $x_3$ from the interface, which do not depend on the amplitude $V_\mathrm{max}$ and frequency $f$ of the electrical excitation. 

\begin{figure*}[!httb]
\centering
\includegraphics[width=0.9\linewidth]{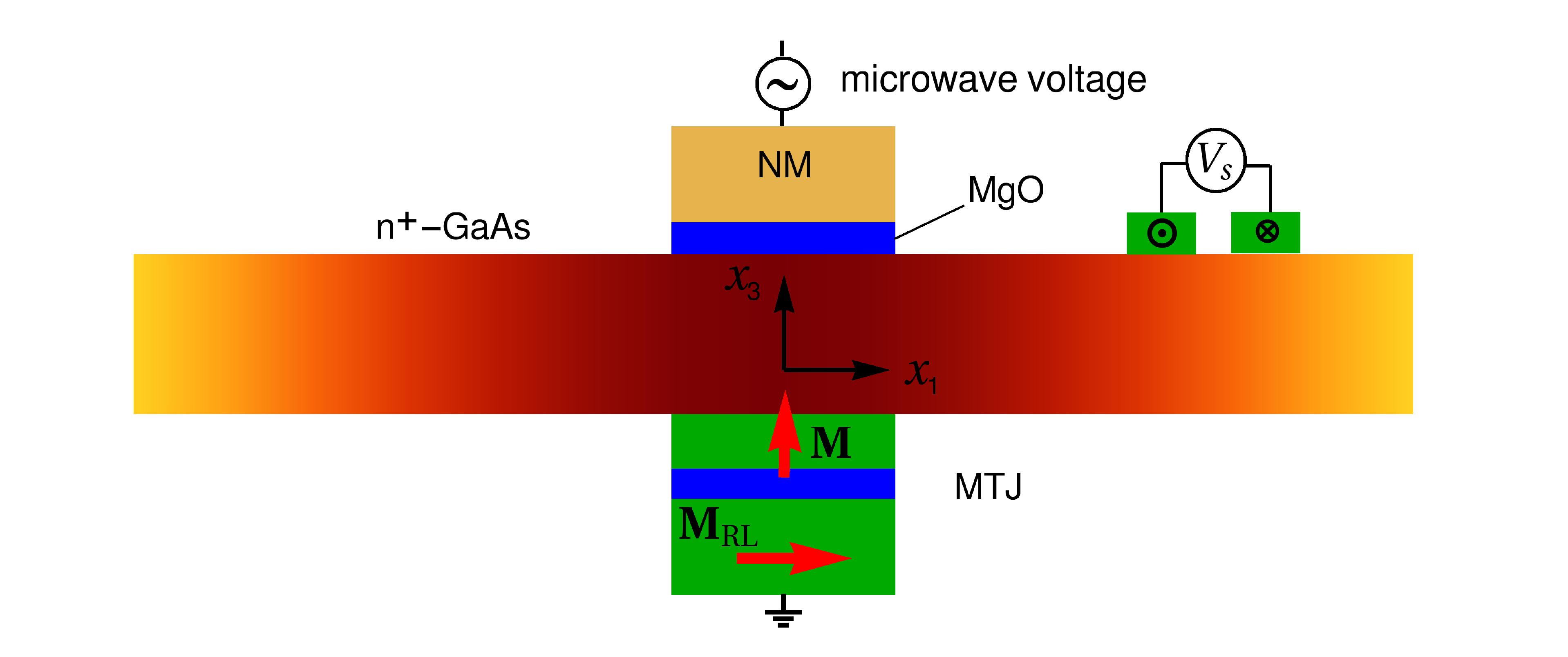}
\caption{\label{fig:SC} Semiconducting GaAs bar sandwiched between electrically excited CoFeB/MgO/CoFeB tunnel junction and MgO/metal bilayer. Information on the spin accumulation in GaAs can be obtained by measuring the voltage $V_s$ between ferromagnetic nanocontacts with antiparallel magnetizations coupled to the bar surface. }
\end{figure*}

For the considered spin injector, the numerical calculations demonstrate that the variations of $\Delta V_1(x_3) / \Delta J_{32}^s$ and $\langle V_1 \rangle(x_3) / \langle J_{32}^s \rangle$ follow the curves shown in Fig.~\ref{fig:Vtr(x3)}. Thus, using the curves in Fig.~\ref{fig:Vtr(x3)} together with the data presented in Fig.~\ref{fig:Js32(Vmax)}, one can evaluate the voltage swing $\Delta V_1(f_\mathrm{res})$ induced by the spin pumping and the time-averaged transverse voltage $\langle V_1 \rangle(f_\mathrm{res})$ proportional to the spin injection. At $V_\mathrm{max} = 600$~mV and $f_\mathrm{res} = 1.16$~GHz, the calculation yields $\Delta V_1 > 6.3$~nV and $\langle V_1 \rangle > 3.6$~nV for the Au region $x_3 < 30$~nm near the interface. The predicted transverse voltages can be detected experimentally, which indicates significant efficiency of the proposed spin injector.

\section{\label{sec:into:SC}SPIN PUMPING INTO SEMICONDUCTOR}

To demonstrate that MTJs excited by microwave voltages can be employed as efficient spin injectors into semiconductors too, we study the spin accumulation in a long GaAs bar connected to the CoFeB/MgO/CoFeB junction and separated from a normal-metal (NM) lead by a thin MgO interlayer (see Fig.~\ref{fig:SC}). The length $L_\mathrm{sc}$ of the GaAs bar is much larger than the FL size $L_1$ along the $x_1$ axis, while its width along the $x_2$ axis is equal to the FL size $L_2$. The considered Si-doped GaAs with the donor concentration $N_D = 10^{18}$~cm$^{-3}$ is a degenerate semiconductor~\cite{Takahashi2003}, which forms an Ohmic contact with CoFeB. Indeed, n$^+$-GaAs has the electron mobility $\chi_\mathrm{sc} = 0.23$~m$^2$~V$^{-1}$~s$^{-1}$~\cite{Kikkawa1998} and the conductivity $\sigma_\mathrm{sc} = \chi_\mathrm{sc} N_D e = 3.68 \times 10^4$~S~m$^{-1}$, which is only one order of magnitude smaller than the conductivity $\sigma_\mathrm{FL} = 4.45 \times 10^5$~S~m$^{-1}$ of CoFeB~\cite{Fan2014}. From the measured spin-flip relaxation time $\tau_\mathrm{sf} = 0.9$~ns~\cite{Bhat2014} and the diffusion coefficient $D = 6 \times 10^{-3}$~m$^2$~s$^{-1}$ obtained via the Einstein relation it follows that the spin-diffusion length $\lambda_\mathrm{sc} = \sqrt{D \tau_\mathrm{sf}}$ in n$^+$-GaAs amounts to about 2.32~$\mu$m at room temperature. Hence the spin accumulation $\bm{\mu}_s(\mathbf{r})$, which is the difference between the chemical potentials for spins parallel and antiparallel to the direction determined by the total nonequilibrium spin-imbalance density~\cite{Tserkovnyak2005}, should be homogeneous in the considered GaAs bar with the nanoscale thickness $t_\mathrm{sc} = 30$~nm $ << \lambda_\mathrm{sd}$ along the $x_3$ axis normal to the CoFeB$|$GaAs interface. Assuming $\bm{\mu}_s(\mathbf{r})$ to be uniform along the $x_2$ axis as well, we obtain a one-dimensional diffusion equation ($k_B$ is the Boltzmann constant)

\begin{equation}
\frac{\partial \bm{\mu}_s}{\partial t} = \frac{4 k_B T}{\hbar N_D t_\mathrm{sc}}\mathbf{e}_n \cdot  \mathbf{J}_\Sigma + D \frac{\partial^2 \bm{\mu}_s}{\partial x_1^2} - \frac{\bm{\mu}_s}{\tau_\mathrm{sf}}
    \label{eq:dmudt}
\end{equation}
\noindent
for the sought function $\bm{\mu}_s(x_1)$. The first term on the r. h. s. of Eq.~(\ref{eq:dmudt}) differs from zero only at $-L_1/2 \leq x_1 \leq L_1/2$ and describes the spin generation in the bar section adjacent to FL. The total spin-current density $\mathbf{J}_\Sigma$ is the sum of four contributions, which result from the spin pumping into GaAs ($\mathbf{J}_\mathrm{sp}$), spin injection from FL into GaAs ($\mathbf{J}_\mathrm{si}$), spin backflow from GaAs to FL ($\mathbf{J}_\mathrm{bf}$), and spin loss caused by the spin-polarized tunnel current flowing across the MgO interlayer separating GaAs from the NM lead ($\mathbf{J}_\mathrm{sl}$). The spin-injection density $\mathbf{J}_\mathrm{si}$ can be calculated using Eq.~(\ref{eq:si}), where the FL spin polarization $p_\mathrm{FL}$ should be replaced by the effective polarization $p_\mathrm{eff} = p_\mathrm{FL}\big[1+(1-p_\mathrm{FL}^2)(\sigma_\mathrm{FL}\lambda_\mathrm{sc})/(\sigma_\mathrm{sc}\lambda_\mathrm{FL})\big]^{-1}$~\cite{Schmidt2000} depending on the CoFeB spin-diffusion length $\lambda_\mathrm{FL}=6.2$~nm~\cite{Zahnd2018}. With the numerical values of the involved parameters the calculation gives very small effective polarization $p_\mathrm{eff} = 1.6 \times 10^{-4}$, which means that the spin injection into n$^+$-GaAs is negligible due to small product $\sigma_\mathrm{sc}\lambda_\mathrm{FL}$ in comparison with $\sigma_\mathrm{FM}\lambda_\mathrm{sc}$. Accordingly, the spin-accumulation vector $\bm{\mu}_s$ appears to be almost orthogonal to the FL magnetization \textbf{M}, and the spin backflow from GaAs to FL can be evaluated via the relation $\mathbf{e}_n \cdot \mathbf{J}_\mathrm{bf} \simeq - \mathrm{Re}\big[g^r_{\uparrow \downarrow}\big]\bm{\mu}_s / 4 \pi$~\cite{Tserkovnyak2002}. Finally, the spin loss $\mathbf{J}_\mathrm{sl}$ caused by the charge current $J_c$ flowing across the GaAs$|$MgO interface equals $\mathbf{e}_n \cdot \mathbf{J}_\mathrm{sl} = -(\hbar / 2e)J_c \bm{\mu}_s / (2 k_B T)$. The distribution $\bm{\mu}_s(x_1)$ of spin accumulation along the GaAs bar was calculated by solving Eq.~(\ref{eq:dmudt}) numerically with the boundary condition $\partial \bm{\mu}_s / \partial x_1 = 0$ at $x_1 = \pm L_\mathrm{sc} / 2$. We assumed that the MgO tunnel barrier separating GaAs from the NM lead has the same conductance $G_\mathrm{P}$ as the CoFeB/MgO/CoFeB junction in the state with parallel electrode magnetizations. Since the GaAs resistance is negligible in comparison with that of two MgO barriers, the dependence of the charge current $J_c$ on voltage $V$ applied to the whole heterostructure was approximated by the relation $J_c = V G_\mathrm{P} G(m_1) / [G_\mathrm{P} + G(m_1)]$. The electrically induced dynamics of the FL magnetization was recalculated with the account of the modified $\tau_\mathrm{STT} = (\gamma \hbar / 2 e) (V_\mathrm{MTJ} G_\mathrm{P} / t_\mathrm{FL}) \eta / (1 + \eta^2)$ and VCMA $K_s = K_s^0 + k_s V_\mathrm{MTJ} / t_b$ resulting from a lower voltage $V_\mathrm{MTJ} = V G_\mathrm{P} / [G_\mathrm{P} + G(m_1)]$ applied to the MTJ. The spin-pumping contribution $\mathbf{J}_\mathrm{sp}$ to the total spin-current density $\mathbf{J}_\Sigma$ involved in Eq.~(\ref{eq:dmudt}) was evaluated using the spin-mixing conductance $\mathrm{Re}\big[ g^r_{\uparrow \downarrow} \big] = 1.5 \times 10^{17}$~m$^{-2}$ determined experimentally for the Ni$_{81}$Fe$_{19}|$GaAs interface~\cite{Ando2011NM}. 

\begin{figure}[b]
\centering
\begin{minipage}[h]{0.8\linewidth}
\includegraphics[width=1.0\linewidth]{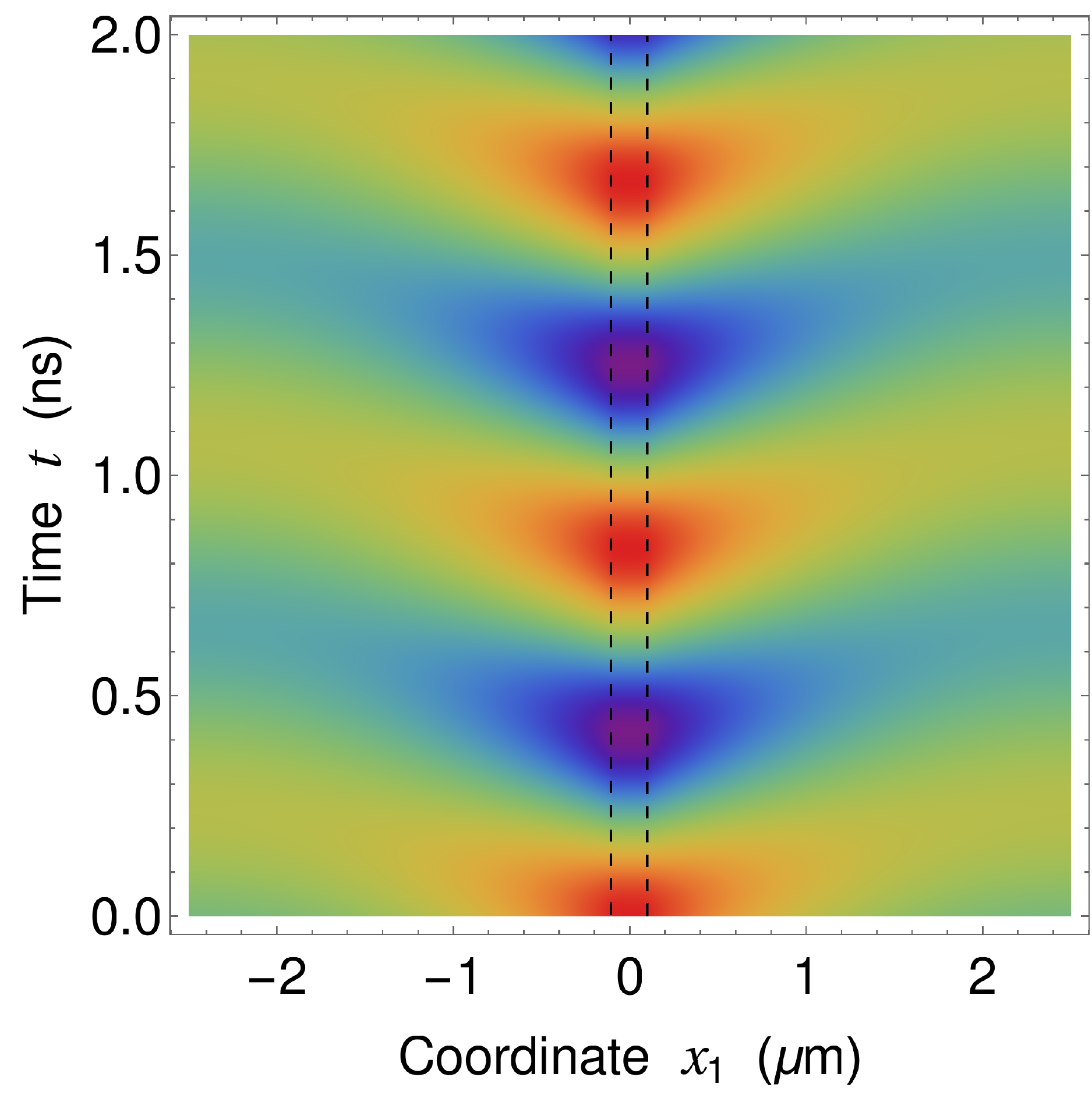}
\end{minipage}
\begin{minipage}[h]{0.12\linewidth}
\includegraphics[width=0.9\linewidth]{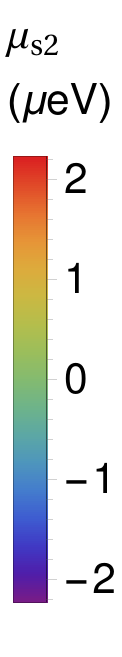}
\end{minipage}
\caption{\label{fig:Mu(x1,t)} Spatio-temporal map of the spin accumulation in the 5-$\mu$m-long n$^+$-GaAs bar coupled to the electrically excited CoFeB/MgO/CoFeB tunnel junction. The map gives the spin-accumulation component $\mu_2^s(x_1, t)$ generated by the microwave voltage with the amplitude $V_\mathrm{max} = 800$~mV and frequency $f = 1.2$~GHz. Vertical dashed lines indicate the surface region, where the bar is coupled to the junction's free layer. }
\end{figure}

\begin{figure*}[!httb]
\centering
\begin{minipage}[h]{0.4\linewidth}
\includegraphics[width=1\linewidth]{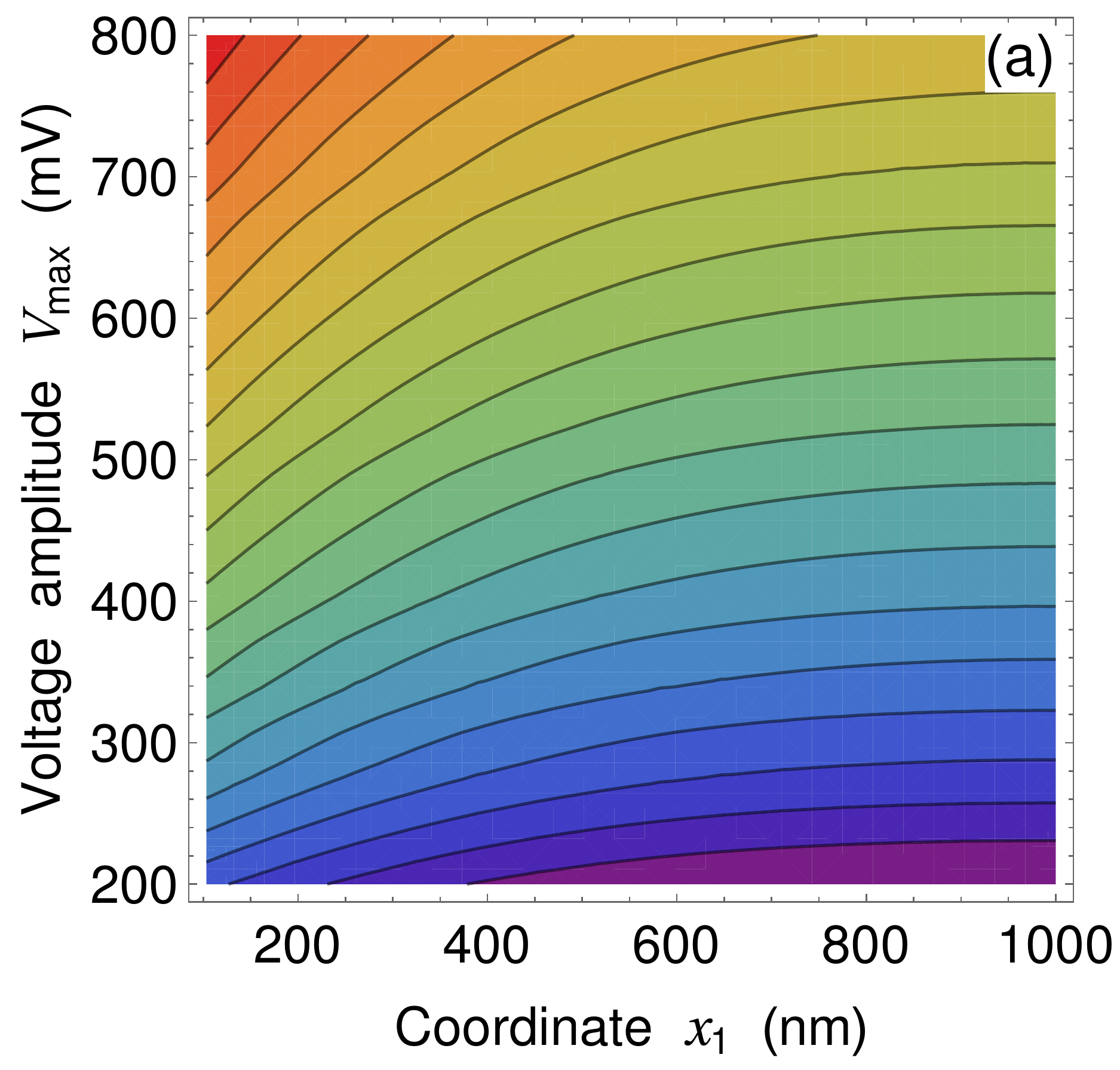}
\end{minipage}
\begin{minipage}[h]{0.08\linewidth}
\includegraphics[width=1\linewidth]{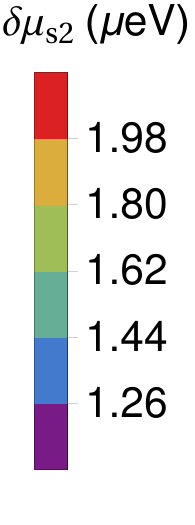}
\end{minipage}
\hfill
\begin{minipage}[h]{0.4\linewidth}
\includegraphics[width=1\linewidth]{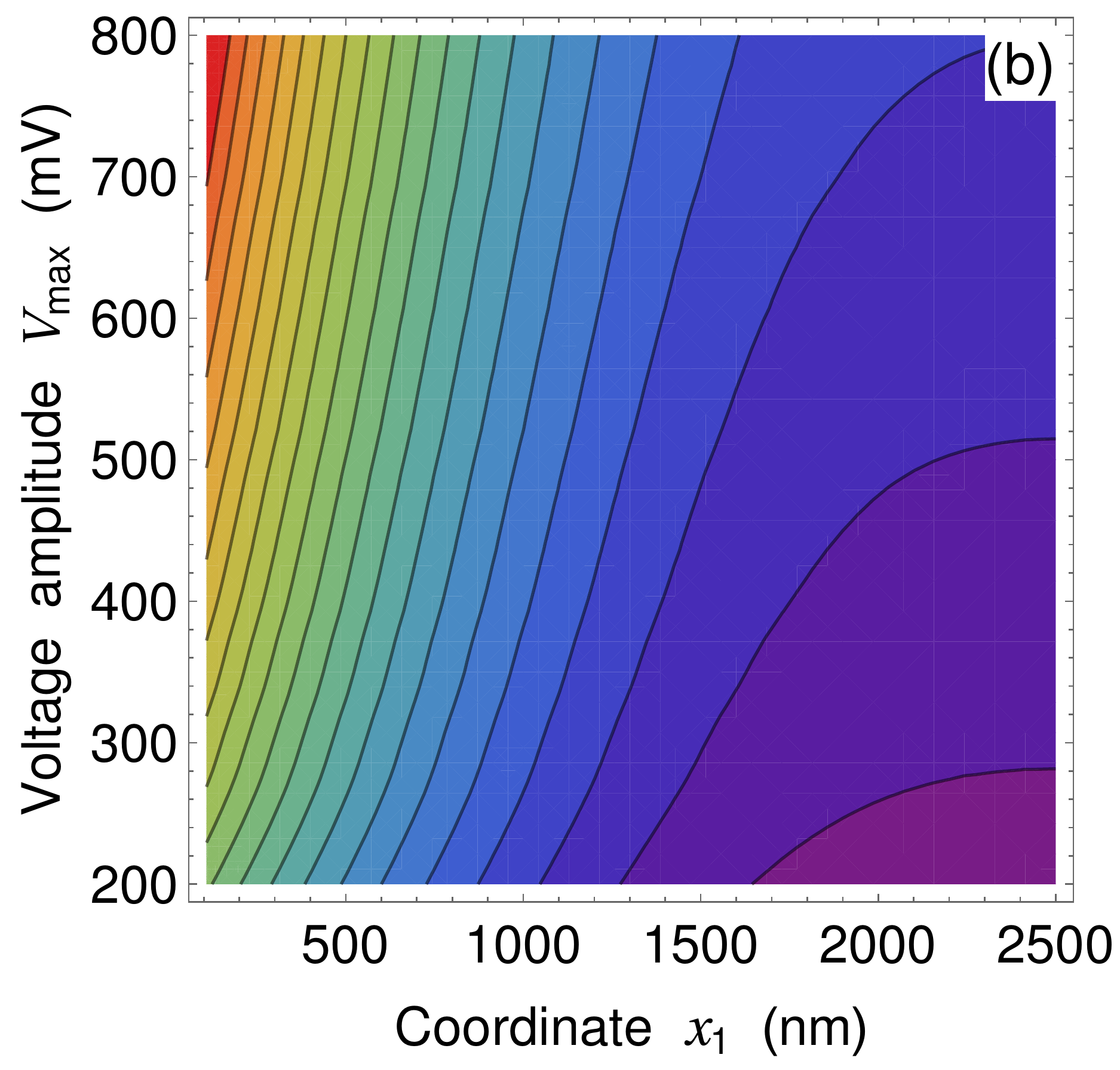}
\end{minipage}
\begin{minipage}[h]{0.08\linewidth}
\includegraphics[width=1\linewidth]{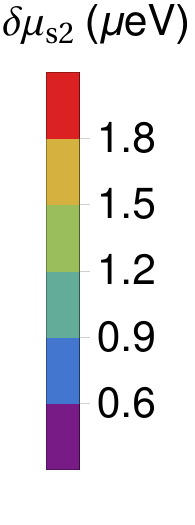}
\end{minipage}
\caption{\label{fig:Mu(x,V)} Spin accumulation in the 2-$\mu$m-long (a) and 5-$\mu$m-long (b) n$^+$-GaAs bars as a function of the spatial position $x_1$ and the amplitude $V_\mathrm{max}$ of the microwave voltage applied to the heterostructure. The maps present the amplitude $\delta \mu_2^s(x_1)$ of the oscillating spin-accumulation component $\mu_2^s(x_1, t)$ calculated at the voltage-dependent resonance frequency $f_\mathrm{res}(V_\mathrm{max})$. Lines show curves on which $\delta \mu_2^s$ remains constant.}
\end{figure*}

The numerical calculations showed that the spin accumulation $\bm{\mu}_s$ in GaAs is determined by the competition of spin pumping $\mathbf{J}_\mathrm{sp}$ and spin backflow $\mathbf{J}_\mathrm{bf}$ across the CoFeB$|$GaAs interface, while the spin loss $\mathbf{J}_\mathrm{sl}$ through the GaAs$|$MgO interface is negligible in comparison with $\mathbf{J}_\mathrm{bf}$. Further, the differences between nonzero components of $\mathbf{J}_\mathrm{sp}$ and $\mathbf{J}_\mathrm{bf}+\mathbf{J}_\mathrm{sl}$ are about 1\% only, which shows that near the CoFeB$|$GaAs interface the spin accumulation is close to saturation. Taking into account that only $\mathbf{J}_\mathrm{sp}$ and $\mathbf{J}_\mathrm{bf}$ create significant contributions to $\mathbf{J}_\Sigma$, we solved Eq.~(\ref{eq:dmudt}) analytically. Since the spin-flip relaxation time $\tau_\mathrm{sf} = 0.9$~ns is comparable to the period $1/ f_\mathrm{res} \sim 1$~ns of magnetization precession, the quasistatic approximation cannot be employed, and $\bm{\mu}_s(x_1)$ should be regarded as a complex quantity. After some mathematical operations, we obtained the following relation between the Fourier components of $\bm{\mu}_s$ and $\mathbf{J}_\mathrm{sp}$:

\begin{equation}
    \begin{gathered}
\pmb{\mu}_s^\omega(x_1)= \frac{4 k_B T \tau_\mathrm{sf}}{\hbar N_D t_\mathrm{sc} (1+ i \omega \tau_\mathrm{sf}) \chi} \mathbf{e}_n \cdot \mathbf{J}^\omega_\mathrm{sp} \\
\times
    \begin{cases}
    \displaystyle\frac{1}{\chi} - \displaystyle\frac{\cosh{(2 x_1 \kappa \chi)}\csch{(L_1 \chi \kappa)}}{\chi^2\coth{[(L_\mathrm{sc}-L_1)\kappa]}+\chi \coth{(L_1 \chi \kappa)}}, & |x_1| < \displaystyle \frac{L_1}{2} \\
    \\
    \displaystyle\frac{\cosh{[(L_\mathrm{sc}-2|x_1|)\kappa]} \csch{[(L_\mathrm{sc}-L_1)\kappa]}}{\chi \coth{[(L_\mathrm{sc}-L_1)\kappa]}+ \coth{(L_1 \chi \kappa)}}, & |x_1| > \displaystyle \frac{L_1}{2},
    \end{cases}
    \end{gathered}
    \label{eq:Mu(x)}
\end{equation}
\noindent
where $\kappa = \sqrt{1 + i \omega \tau_\mathrm{sf}} / (2 \lambda_\mathrm{sc})$ and $\chi = \sqrt{1 + k_B T \tau_\mathrm{sf} \mathrm{Re}\big[ g^r_{\uparrow \downarrow} \big] / [\pi \hbar N_D t_\mathrm{sc} (1+i \omega \tau_\mathrm{sf})]}$. By combining Eq.~(\ref{eq:Mu(x)}) with numerical results obtained for the electrically driven precession of the FL magnetization $\mathbf{m}(t)$ and the accompanying spin pumping $\mathbf{J}_\mathrm{sp}(\mathbf{m})$, one can calculate the spin accumulation $\bm{\mu}_s$ as a function of the distance $|x_1|$ from the bar center and the time $t$. At small frequencies $\omega << 1 / \tau_\mathrm{sf}$, the parameter $\kappa$ is real, and the phase of $\bm{\mu}_s(t)$ does not depend on the position $x_1$. However, at the precession frequencies $f \sim 1$~GHz the spin accumulation $\bm{\mu}_s(x_1,t)$ has a position-dependent delay from the applied voltage $V(t)$.

Figure~\ref{fig:Mu(x1,t)} shows the spatio-temporal map of the spin-accumulation component $\mu_2^s(x_1,t)$ generated in 5-$\mu$m-long n$^+$-GaAs bar by the applied voltage with amplitude $V_\mathrm{max} = 800$~mV and frequency $f = 1.2$~GHz. Since this frequency corresponds to the main peak of the magnetization precession, we assume that $J_{32}^\mathrm{sp}$ and $\mu_2$ exhibit almost simple harmonic oscillations with the excitation frequency $f$. However, these oscillations lag behind the applied microwave voltage by about 0.25~ns at $|x_1| = L_\mathrm{sc} / 2$ in the 5-$\mu$m-long bar (see Fig.~\ref{fig:Mu(x1,t)}).

The averaging of $\mu_2^s(x_1,t)$ over the oscillation period shows that the mean spin accumulation $\langle \mu_2^s(x_1) \rangle$ is negligible in comparison with the oscillation amplitude $\delta \mu_2^s(x_1)$. The maps presented in Fig.~\ref{fig:Mu(x,V)} demonstrate how the magnitude $V_\mathrm{max}$ of applied microwave voltage influences the spatial distribution of $\delta \mu_2^s(x_1)$ calculated at the voltage-dependent resonance frequency $f_\mathrm{res}(V_\mathrm{max})$. Remarkably, the ac component of the spin accumulation remains significant even at the ends of the considered GaAs bars with the length $L_\mathrm{sc}$ ranging from 2 to 5~$\mu$m. When $L_\mathrm{sc}$ is smaller than the spin diffusion length $\lambda_\mathrm{sc} = 2.32$~$\mu$m, $\delta \mu_2^s(x_1)$ appears to be weakly dependent on the coordinate $x_1$ near the bar ends [Fig.~\ref{fig:Mu(x,V)}(a)]. In contrast, it decreases more rapidly with the distance from the bar center at $L_\mathrm{sc} > \lambda_\mathrm{sc}$~[Fig.~\ref{fig:Mu(x,V)}(b)] and becomes smaller everywhere due to the spreading of nonequilibrium spin imbalance in a larger volume.

The spin accumulation in the GaAs bar can be measured experimentally by a method similar to a nonlocal detection of the spin injection into a normal conductor~\cite{JohnsonSilsbee1985, Lou2007}. The method employs two ferromagnetic nanostrips integrated onto the bar surface and connected to a voltmeter (see Fig.~\ref{fig:SC}). The nanostrips should be oriented along the $x_2$ axis and have antiparallel in-plane magnetizations. The presence of the spin accumulation $\mu_2^s(x_1,t)$ in the GaAs region beneath ferromagnetic strips with nanoscale widths $\sim 10$~nm and small separation $\Delta x_1 << x_1$ manifests itself in a voltage $V_s(x_1, t) \propto \mu_2^s(x_1, t)$ between the nanocontacts. In the case of Fe nanocontacts forming Schottky tunnel barriers with n$^+$-GaAs, $V_s(x_1, t) = \eta_\mathrm{IE} p_\mathrm{Fe} \mu_2^s(x_1, t) / e$, where $\eta_\mathrm{IE} \approx 0.5$ is the spin transmission efficiency of the GaAs$|$Fe interface and $p_\mathrm{Fe} \approx 0.42$ is the spin polarization of Fe at the Fermi level~\cite{Lou2007}. Using this relation, we calculated frequency spectra of the spin signals $V_s(x_1, t)$ generated in the 0.8-$\mu$m-long n$^+$-GaAs bar at different excitation frequencies $f$. 
\begin{figure}[b]
\centering
\begin{minipage}[h]{0.8\linewidth}
\includegraphics[width=1.0\linewidth]{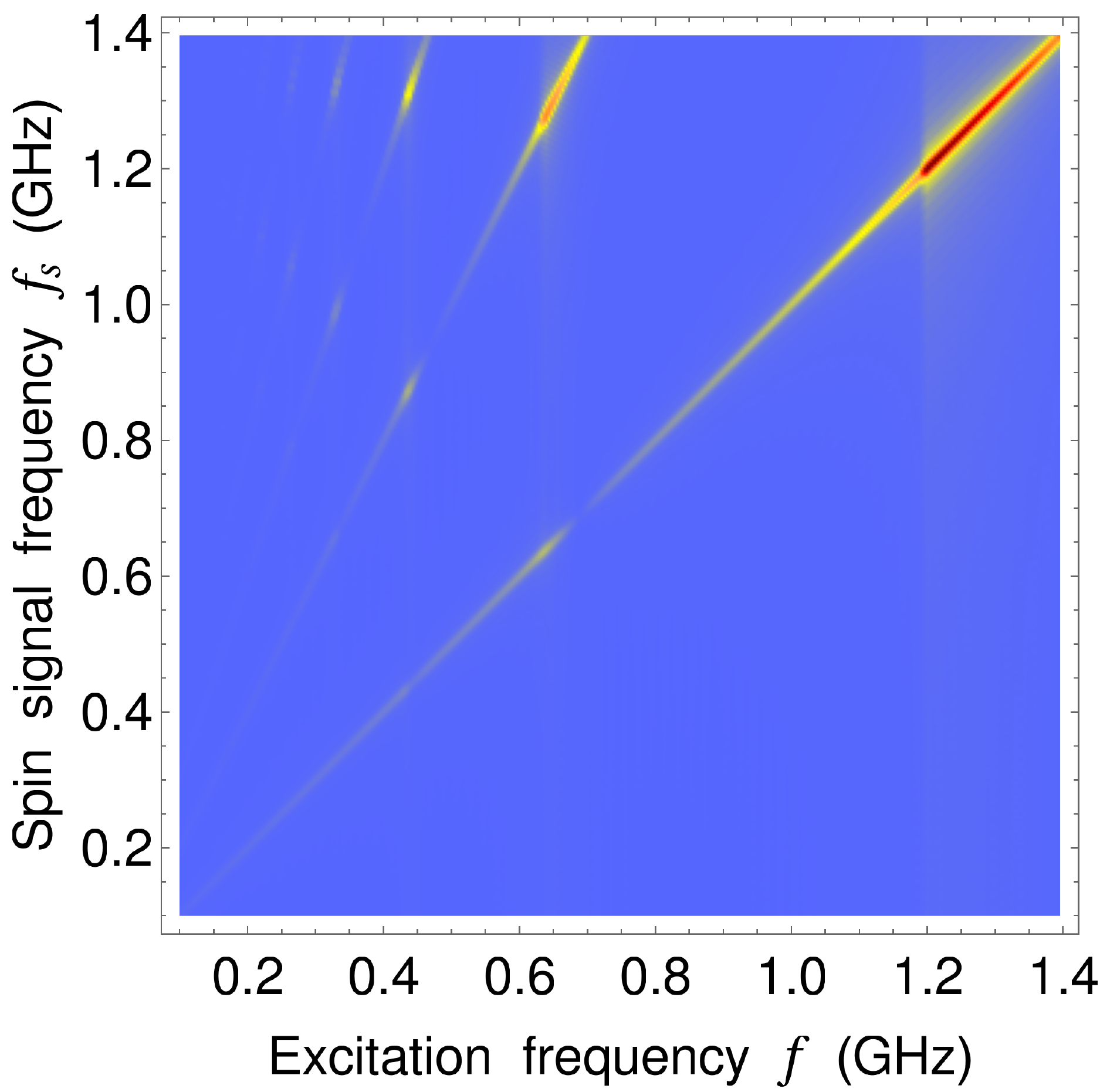}
\end{minipage}
\begin{minipage}[h]{0.14\linewidth}
\includegraphics[width=0.9\linewidth]{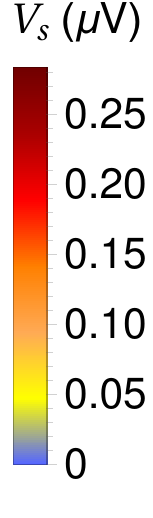}
\end{minipage}
\caption{\label{fig:spectra} Frequency spectra of the spin signals $V_s(t)$ generated in the 0.8-$\mu$m-long n$^+$-GaAs bar at different excitation frequencies $f$. The map shows amplitudes of the Fourier components of $V_s(t)$ at the spatial position $x_1 = 300$~nm. The applied voltage has the amplitude $V_\mathrm{max} = 800$~mV, and its frequency increases from lower to higher values. }
\end{figure}
The map presented in Fig.~\ref{fig:spectra} demonstrates amplitudes of the Fourier components of $V_s(t)$ determined at $V_\mathrm{max} = 800$~mV and the distance $|x_1| = 300 $~nm from the bar center. It can be seen that the maximal ac spin signal with the amplitude of about 0.26~$\mu$V and frequency $f_s = f$ appears at the excitation frequencies $f = 1.2-1.4$~GHz close to the resonance frequency $f_\mathrm{res}$. 
\begin{figure}[!h]
\centering
\includegraphics[width=0.8\linewidth]{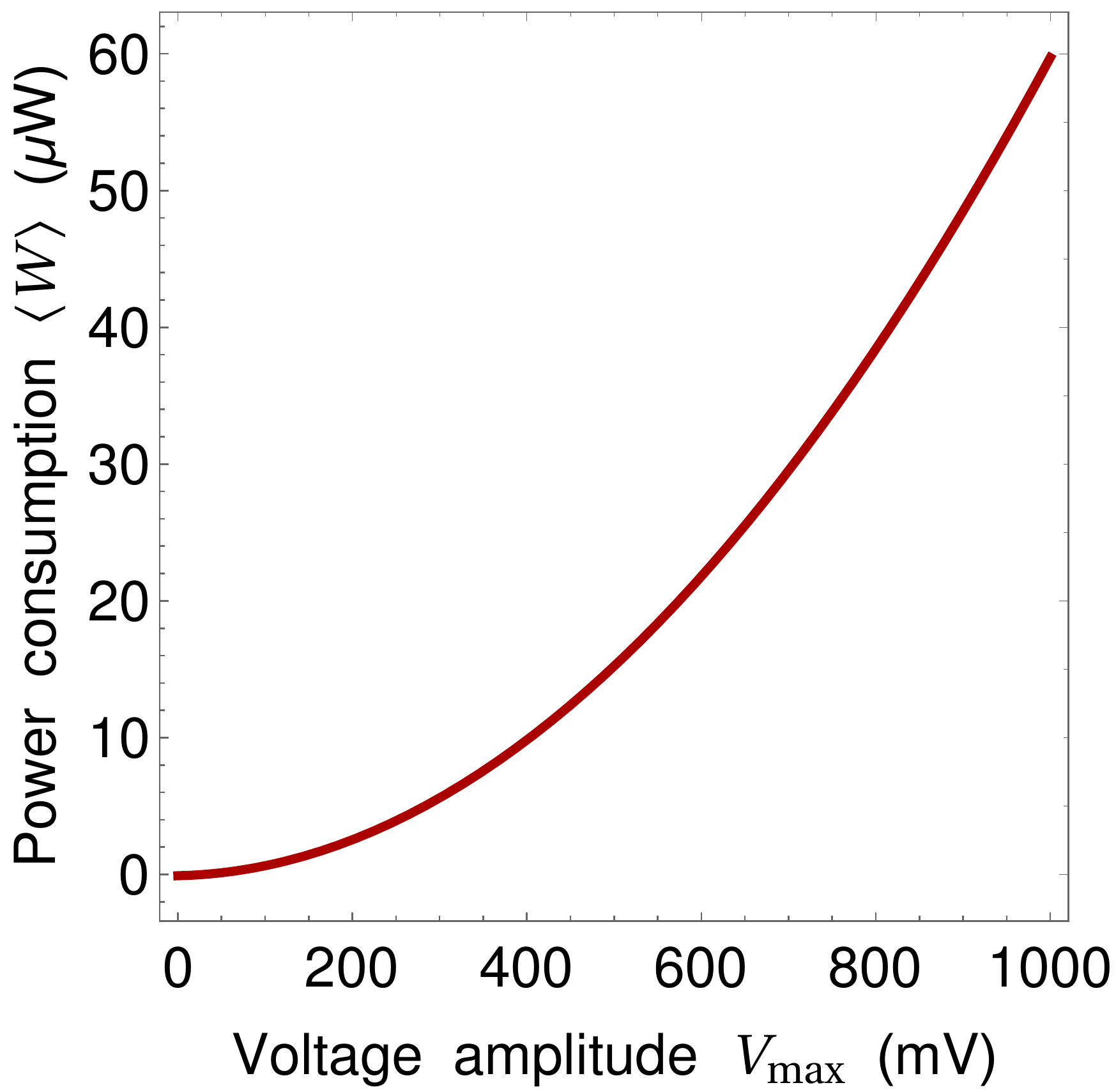}
\caption{\label{fig:power} Power consumption of the CoFeB/MgO/CoFeB/GaAs/ MgO heterostructure as a function of the amplitude of the applied microwave voltage. }
\end{figure}
In addition, the Fourier components of $V_s(t)$ with frequencies $f_s$ slightly above $f_\mathrm{res}$ exhibit smaller maxima at the excitation frequencies $f \approx f_\mathrm{res}/n$, which correspond to secondary peaks of the precession amplitude similar to those shown in Fig.~\ref{fig:res}(c). Remarkably, such spin signals have significant amplitudes ranging from 0.12~$\mu$V at $f \approx  f_\mathrm{res}/2$ to 0.04~$\mu$V at $f \approx f_\mathrm{res}/3$, which can be detected experimentally. Thus, the proposed spin injector allows the generation of the ac spin accumulation in GaAs oscillating with the microwave frequency $f_s \approx f_\mathrm{res}$ several times higher than the excitation frequency. Such frequency multiplication could be useful for device applications.

The mean power consumption $\langle W \rangle$ of the proposed spin injector can be estimated from the relation 
\begin{equation}
\langle W \rangle = f L_1 L_2 \int_0^{1/f}{J_c[t,m_1(t)] V_\mathrm{ac}(t)dt},
\label{eq:W}
\end{equation}
where the integral is taken over the period of an applied ac voltage. Figure~\ref{fig:power} shows the dependence $\langle W \rangle(V_\mathrm{max})$ calculated for the CoFeB/MgO/CoFeB/GaAs/MgO heterostructure considered in this work. It can be seen that $\langle W \rangle \propto V_\mathrm{max}^2$ rises rapidly with the voltage amplitude, but remains well below 100~$\mu$W even at $V_\mathrm{max} = 1$~V. Hence the power dissipation of the electrically driven spin injector is more than two orders of magnitude smaller than that of the device excited by the microwave magnetic field ($\sim 10$~mW)~\cite{Ando2011NM}.

\section{\label{sec:intro}CONCLUSIONS}

In this paper, we theoretically described the spin dynamics in the Co$_{20}$Fe$_{60}$B$_{20}$/MgO/Co$_{20}$Fe$_{60}$B$_{20}$/Au and Co$_{20}$Fe$_{60}$B$_{20}$/MgO/Co$_{20}$Fe$_{60}$B$_{20}$/GaAs heterostructures subjected to a microwave voltage. Our calculations were focused on the heterostructures comprising a nanoscale Co$_{20}$Fe$_{60}$B$_{20}$/MgO/Co$_{20}$Fe$_{60}$B$_{20}$ tunnel junction with an ultrathin FL having perpendicular magnetic anisotropy (Fig.~\ref{fig:mtj}). By solving the LLGS equation numerically, we first quantified the electrically induced precession of FL magnetization with the account of STT created by the spin-polarized current flowing through FL, VCMA associated with the FL$|$MgO interface, and enhanced Gilbert damping caused by the spin pumping into the overlayer. The calculated dependences of the precession amplitude on the frequency $f$ and magnitude $V_\mathrm{max}$ of the applied voltage showed that FL exhibits strongly nonlinear dynamic behavior at $V_\mathrm{max} > 200$~mV. In particular, the main peak of precession amplitude located at the resonance frequency $f_\mathrm{res}$ becomes asymmetric with a break on the left side [Fig.~\ref{fig:res}(b)], which is similar to the behavior of a Duffing oscillator with a softening nonlinearity~\cite{Nayfeh1979}. At higher applied voltages $V_\mathrm{max} > 500$~mV, the frequency dependence of precession amplitude also involves strong secondary peaks and additional breaks [Fig.~\ref{fig:res}(c)].

The description of the magnetization dynamics occuring in the Co$_{20}$Fe$_{60}$B$_{20}$ FL enabled us to quantify the spin injection and pumping into the Au and GaAs overlayers. The total spin-current densities $J^s_{3i}$ generated near the interface were calculated as a function of time at different frequencies and amplitudes of the applied voltage (Fig.~\ref{fig:SC(t)}). The analysis of these time dependences showed that the densities $J^s_{3i}(t)$ contain both ac and dc components, which mostly maximize under resonant excitation (Figs.~\ref{fig:AC(f)} and~\ref{fig:DC(f)}). Interestingly, the ac components $\Delta J^s_{31}$ and $\Delta J^s_{32}$ also strongly increase at the excitation frequencies $f \approx f_\mathrm{res}/2$ and high voltage amplitudes $V_\mathrm{max} \geq 400$~mV.

To evaluate the efficiency of spin generation in Au, we determined the distribution of electric potential in the 200-nm-thick Au overlayer by solving the Laplace's equation. The charge current flowing in the overlayer was calculated with the account of the drift contribution and the inverse spin Hall effect. When finding the spatial distribution of the actual spin-current density $\mathbf{J}_\mathrm{Au}$ we considered the spin injection and pumping at the FL$|$Au interface, spin relaxation and diffusion inside Au, and the spin backflow into FL. The calculated distribution of the electric potential $\phi$ was used to determine the transverse voltage $V_1(x_3) = \phi(x_1=L_1,x_3) - \phi(x_1=0,x_3)$ between the sides of Au overlayer normal to the $x_1$ axis parallel to the RL magnetization. It was found that both ac and dc components of this time-dependent voltage can be measured experimentally at small distances $x_3 < 30$~nm from the FL$|$Au interface under excitation by the microwave voltage with $f = f_\mathrm{res}$ and $V_\mathrm{max} = 600$~mV. The measured dependence $V_1(x_3)$ provides information on the spatial decay of the actual spin-current density $J_{32}^\mathrm{Au}$ reduced by spin backflow into FL and spin relaxation in Au.

In the final part of this study, we quantified the time-dependent spin accumulation in the n$^+$-GaAs bar coupled to the CoFeB/MgO/CoFeB junction at the center and separated from the NM lead by a thin MgO interlayer (Fig.~\ref{fig:SC}). By solving numerically the spin diffusion equation with appropriate boundary conditions, we calculated the spatio-temporal map of the spin-accumulation component $\mu_2^s(x_1,t)$ in the bar (Fig.~\ref{fig:Mu(x1,t)}). It was revealed that there is a position-dependent delay of $\mu_2^s(x_1,t)$ from the applied microwave voltage $V_\mathrm{ac}(t)$. The time-averaged value $\langle \mu_2^s(x_1) \rangle$ of the oscillating spin accumulation was found to be negligible in comparison with the oscillation amplitude $\delta \mu_2^s(x_1)$. At the same time, the ac component $\delta \mu_2^s(x_1)$ of the spin accumulation under resonant excitation remained to be significant even at the ends of the 5-$\mu$m-long n$^+$-GaAs bar. To detect this component, we proposed to use two ferromagnetic nanostrips integrated onto the bar surface and connected to a voltmeter (Fig.~\ref{fig:SC}). The voltage $V_s(x_1,t) \propto \mu_2^s(x_1,t)$ between such nanocontacts and its frequency spectrum were calculated. The results showed that the maximal ac spin signal appears at the excitation frequencies $f=1.2-1.4$~GHz close to the resonance frequency $f_\mathrm{res}$. Its frequency $f_s$ is equal to the excitation one, and the amplitude is about 0.26~$\mu$V at a representative distance $|x_1| = 300$~nm from the bar center. In addition, the Fourier components of $V_s(t)$ with frequencies $f_s$ slightly above $f_\mathrm{res}$ exhibit significant maxima at the excitation frequencies about $f_\mathrm{res}/2$ and $f_\mathrm{res}/3$. These results demonstrate high efficiency of the described nanoscale spin injector and the possibility of ac spin accumulation with frequency multiplication. It should be noted that the proposed device is distinguished from the spin injector driven by a microwave magnetic field~\cite{Ando2011NM} by a compact design and low power consumption.

\begin{acknowledgments}
This work was supported by the Foundation for the Advancement of Theoretical Physics and Mathematics ``BASIS''.
\end{acknowledgments}

\bibliography{apssamp}

\end{document}